\documentclass[longauth]{aa} 

\usepackage{color}
\usepackage{graphicx}
\usepackage{txfonts}
\usepackage{hyperref}
\usepackage{natbib}
\bibpunct{(}{)}{;}{a}{}{,}             
\begin{document} 

\title{The \emph{Gaia}-ESO Survey: Sodium and aluminium abundances in giants and dwarfs\thanks{Based on observations made with the ESO/VLT, at Paranal Observatory, under program 188.B-3002 (The Gaia-ESO Public Spectroscopic Survey), and on data obtained from the ESO Archive originally observed under programs 60.A-9143, 076.B-0263 and 082.D-0726.}\fnmsep\thanks{Table 1 is only available in electronic form at the CDS via anonymous ftp to cdsarc.u-strasbg.fr (130.79.128.5) or via http://cdsweb.u-strasbg.fr/cgi-bin/qcat?J/A+A/}}

\subtitle{Implications for stellar and Galactic chemical evolution}

\author{R. Smiljanic\inst{1}
         \and
         D. Romano\inst{2}
          \and
         A. Bragaglia\inst{2}          
          \and
         P. Donati\inst{2,3}          
          \and
         L. Magrini\inst{4}
         \and
         E. Friel\inst{5}         
         \and
         H. Jacobson\inst{6}
         \and
         S. Randich\inst{4}
          \and
          P. Ventura\inst{7}
          \and
          K. Lind\inst{8,9}
          \and
          M. Bergemann\inst{9}
          \and
          T. Nordlander\inst{8}
          \and
         T. Morel\inst{10}
          \and
          E. Pancino\inst{2,11}
          \and
          G. Tautvai{\v s}ien{\.e}\inst{12}
         \and
         V. Adibekyan\inst{13}
         \and
         M. Tosi\inst{2}
         \and
         A. Vallenari\inst{14}
         \and
         G. Gilmore\inst{15}
         \and
         T. Bensby\inst{16}
         \and
         P. Fran\c cois\inst{17,18}
         \and
         S. Koposov\inst{15}
         \and
         A.~C. Lanzafame\inst{19}
         \and
         A. Recio-Blanco\inst{20}
         \and
         A. Bayo\inst{21}
         \and
         G. Carraro\inst{22}
         \and
         A.~R. Casey\inst{15}
         \and
         M.~T. Costado\inst{23}
         \and
         E. Franciosini\inst{4}
         \and
         U. Heiter\inst{24}
         \and
         V. Hill\inst{20}
         \and
         A. Hourihane\inst{15}
         \and
         P. Jofr\'e\inst{15}
         \and
         C. Lardo\inst{25}
         \and
         P. de Laverny\inst{20}
         \and
         J. Lewis\inst{15}
         \and
         L. Monaco\inst{26}
         \and
         L. Morbidelli\inst{4}
         \and
         G.~.G. Sacco\inst{4}
         \and
         L. Sbordone\inst{27,28}
         \and
         S.~G. Sousa\inst{13}
         \and
         C.~C. Worley\inst{15}
         \and
         S. Zaggia\inst{14}
          }

\institute{
           Department for Astrophysics, Nicolaus Copernicus Astronomical Center, 
           ul. Rabia\'nska 8, 87-100 Toru\'n, Poland \\
           \email{rsmiljanic@ncac.torun.pl}
            \and
           INAF-Osservatorio Astronomico di Bologna, Via Ranzani 1, I-40127 Bologna, Italy
           \and
           Dipartimento di Fisica e Astronomia, Universit\`a di Bologna, Via Ranzani 1, I-40127 Bologna, Italy  
           \and
           INAF-Osservatorio Astrofisico di Arcetri, Largo Enrico Fermi 5 50125 Firenze, Italy
           \and
           Department of Astronomy, Indiana University, Bloomington, IN 47405, USA
           \and
            Kavli Institute for Astrophysics \& Space Research, Massachusetts Institute of Technology, 77 Massachusetts Avenue, Cambridge, MA 02139 USA
           \and
           INAF - Osservatorio Astronomico di Roma, Via Frascati 33, I-00040 Monte Porzio Catone (RM), Italy 
           \and
           Department of Physics and Astronomy, Uppsala University, Box 516, SE-751 20 Uppsala, Sweden             
           \and
           Max Planck Institute f\"ur Astronomy, K\"onigstuhl 17, D-69117 Heidelberg, Germany 
           \and
           Institut d'Astrophysique et de G\'eophysique, Universit\'e de Li\`ege, All\'ee du 6 Ao\^ut, B\^at. B5c, 4000 Li\`ege, Belgium        
           \and
         ASI Science Data Center, via del Politecnico SNC, I-00133, Roma, Italy
           \and
         Institute of Theoretical Physics and Astronomy, Vilnius University, Go{\v s}tauto 12, Vilnius LT-01108, Lithuania
           \and
           Instituto de Astrof\'isica e Ci\^encias do Espa\c{c}o, Universidade do Porto, CAUP, Rua das Estrelas, 4150-762 Porto, Portugal
           \and
           INAF - Osservatorio Astronomico di Padova, Vicolo Osservatorio 2 I-35122 Padova, Italy 
           \and
           Institute of Astronomy, University of Cambridge, Madingley Road, Cambridge CB3 0HA, United Kingdom
           \and
           Lund Observatory, Department of Astronomy and Theoretical Physics, Box 43, SE-221 00 Lund, Sweden
           \and
           GEPI, Observatoire de Paris, PSL Research University, CNRS, Univ Paris Diderot, Sorbonne Paris Cit\'e, 61 Avenue de l'Observatoire, 75014 Paris, France
%
           \and 
           Universit\'e de Picardie Jules Verne, Physics Dpt. 33 rue St Leu, F-80000 Amiens, France
           \and
           Dipartimento di Fisica e Astronomia, Sezione Astrofisica, Universit\'a di Catania, via S. Sofia 78, 95123, Catania, Italy
           \and
           Laboratoire Lagrange (UMR7293), Universit\'e de Nice Sophia Antipolis, CNRS,Observatoire de la C\^ote d'Azur, CS 34229,F-06304 Nice cedex 4, France
           \and
           Instituto de F\'{i}sica y Astronom\'{i}a, Universidad de Valpara\'{i}so, Chile
           \and
           European Southern Observatory, Alonso de Cordova 3107 Vitacura, Santiago de Chile, Chile
           \and
           Instituto de Astrof\'{i}sica de Andaluc\'{i}a-CSIC, Apdo. 3004, 18080 Granada, Spain
           \and
           Department of Physics and Astronomy, Uppsala University, Box 516, SE-751 20 Uppsala, Sweden
           \and
           Astrophysics Research Institute, Liverpool John Moores University, 146 Brownlow Hill, Liverpool L3 5RF, United Kingdom
           \and
           Departamento de Ciencias Fisicas, Universidad Andres Bello, Republica 220, Santiago, Chile
           \and
           Millennium Institute of Astrophysics, Av. Vicu\~{n}a Mackenna 4860, 782-0436 Macul, Santiago, Chile
           \and
           Pontificia Universidad Cat\'{o}lica de Chile, Av. Vicu\~{n}a Mackenna 4860, 782-0436 Macul, Santiago, Chile
           }

   \date{Received one day; accepted some time later}

\titlerunning{Stellar and Galactic chemical evolution of Na and Al}
\authorrunning{Smiljanic et al.}

 
  \abstract
{Stellar evolution models predict that internal mixing should cause some sodium overabundance at the surface of red giants more massive than $\sim$ 1.5--2.0 $M_{\odot}$. The surface aluminium abundance should not be affected. Nevertheless, observational results disagree about the presence and/or the degree of Na and Al overabundances. In addition, Galactic chemical evolution models adopting different stellar yields lead to very different predictions for the behavior of [Na/Fe] and [Al/Fe] versus [Fe/H]. Overall, the observed trends of these abundances with metallicity are not well reproduced.}
{We readdress both issues, using new Na and Al abundances determined within the Gaia-ESO Survey. Our aim is to obtain better observational constraints on the behavior of these elements using two samples: i) more than 600 dwarfs of the solar neighborhood and of open clusters and ii) low- and intermediate-mass clump giants in six open clusters.}
{Abundances were determined using high-resolution UVES spectra. The individual Na abundances were corrected for nonlocal thermodynamic equilibrium effects. For the Al abundances, the order of magnitude of the corrections was estimated for a few representative cases. For giants, the abundance trends with stellar mass are compared to stellar evolution models. For dwarfs, the abundance trends with metallicity and age are compared to detailed chemical evolution models. }
{Abundances of Na in stars with mass below $\sim$2.0 $M_{\odot}$, and of Al in stars below $\sim$3.0 $M_{\odot}$, seem to be unaffected by internal mixing processes. For more massive stars, the Na overabundance increases with stellar mass. This trend agrees well with predictions of stellar evolutionary models. For Al, our only cluster with giants more massive than 3.0 M$_{\odot}$, NGC 6705, is Al enriched. However, this might be related to the environment where the cluster was formed. Chemical evolution models that well fit the observed [Na/Fe] vs. [Fe/H] trend in solar neighborhood dwarfs cannot simultaneously explain the run of [Al/Fe] with [Fe/H], and vice versa. The comparison with stellar ages is hampered by severe uncertainties. Indeed, reliable age estimates are available for only a half of the stars of the sample. We conclude that Al is underproduced by the models, except for stellar ages younger than about 7 Gyr. In addition, some significant source of late Na production seems to be missing in the models. Either current Na and Al yields are affected by large uncertainties, and/or some important Galactic source(s) of these elements has as yet not been taken into account.}
  {}

   \maketitle
%
\section{Introduction}

Sodium and aluminium are two odd-Z elements with single stable isotopes ($^{23}$Na and $^{27}$Al, respectively) of importance for studies of stellar and Galactic chemical evolution. In a Galactic context, Na is mainly synthesized during hydrostatic carbon burning in massive stars \citep{1952ApJ...115..326S,1959ApJ...130..429C}, where its final abundance is also sensitive to the neutron excess \citep{1995ApJS..101..181W}. Sodium is also produced in high-temperature H-burning regions through the NeNa cycle \citep{1955PhRv...97.1237S,1990SvAL...16..275D}. In low- and intermediate-mass stars, Na produced by the NeNa cycle can potentially be mixed to the stellar surface either during the first dredge-up or later during the asymptotic giant branch (AGB) phase \citep[see, e.g.,][]{1995ApJ...451..298E,1999A&A...350...73M,2010MNRAS.403.1413K}. Aluminium is mainly synthesized during carbon and neon burning in massive stars \citep{1985ApJ...295..589A,1985ApJ...295..604T}. It can also be produced through the MgAl cycle in the internal convective regions of AGB stars with an initial mass above $\sim$ 5 $M_{\odot}$, which are undergoing hot bottom burning \citep{2013MNRAS.431.3642V,2014MNRAS.437..195D}.

\pagebreak

Abundances of Na and Al have been determined in local disk and halo stars in a number of works \citep[e.g.,][]{1962ApJS....6..407W,1980A&A....89..118S,1981ApJ...244..989P,1986A&A...160..264F,1986A&A...165..183F,1993A&A...275..101E,1995AJ....109.2757M,1996AJ....111.1689P,2000A&A...356..238C,2004A&A...416.1117C,2004A&A...413.1045G,2006AJ....131.3069L,2006MNRAS.367.1329R,2008A&A...489..923M,2012A&A...545A..32A,2014AstL...40..406A,2014A&A...562A..71B}. The observed trends with metallicity are different for the two elements. For sodium, a mean trend of increasing [Na/Fe] for super-solar metallicities is of particular interest. This is not seen for [Al/Fe]. The [Al/Fe] ratio increases with decreasing metallicity up to [Al/Fe] $\sim$ +0.4 at [Fe/H] $\sim$ $-$1.0, where it decreases again. The increase of [Na/Fe] for low metallicities is less pronounced. These trends are discussed in Sect.\ \ref{sec:chemical}.

Chemical evolution models still have problems reproducing the observed behavior of the Na and Al abundances. Depending on the stellar yields adopted by the models, different regions of the [Na/Fe] or [Al/Fe] vs.\ [Fe/H] diagrams can be fit, but a complete explanation of the detailed trends is not achieved \citep[e.g.,][]{2010A&A...522A..32R,2013ARA&A..51..457N}. The increase in [Na/Fe] for super-solar metallicities is a particular challenge. For example, none of the models computed by \citet{2010A&A...522A..32R}, with different stellar yields, was able to reproduce such behavior.

On the stellar evolution side, it is not clear to what extent the first dredge-up \citep{1964ApJ...140.1631I,1967ARA&A...5..571I} in low- and intermediate-mass stars is able to bring the products of the NeNa cycle to the stellar photosphere. Stellar evolution models predict that mixing is deep enough to change the Na abundance only in giants above $\sim$ 1.5--2.0 $M_{\odot}$ \citep[see e.g.,][]{2010A&A...522A..10C}, particularly in those of intermediate-mass above $\sim$ 4.0 $M_{\odot}$ \citep[see e.g.,][]{1995ApJ...451..298E,2005ApJ...622.1058D}. The Al surface abundance is not expected to increase during the giant phase because no magnesium burning is activated in the central region of H-burning of these stars \citep[][]{2004MmSAI..75..347W}.

Observationally, it is well known that evolved intermediate-mass stars show some Na enhancement after the first dredge-up \citep{1994PASJ...46..395T,2002A&A...389..519A,2005PASP..117.1173K,2013MNRAS.432..769T}, although a small excess of Na from Galactic chemical evolution cannot be fully excluded.

For low-mass stars (0.80 $\leq$ M/$M_{\odot}$ $\leq$ 2.5), the situation is more confusing. Low-mass metal-poor field giants do not show an indication of changes in their surface Na abundances \citep{2000A&A...354..169G}. However, approximately 65\% of giants in open clusters (stars with higher metallicity and a wider range of masses) seem to have enhanced Na and/or Al abundances \citep[see e.g.,][and references therein]{2007AJ....134.1216J,2009A&A...502..267S,2010A&A...511A..56P,2011A&A...535A..30C,2012MNRAS.422.1562S,2012AJ....144...95Y}. Sodium and/or aluminium overabundances are sometimes detected in field giants also \citep[see e.g.,][]{2006A&A...456.1109M,2015MNRAS.450.1900A}.

The level of the Na overabundances varies depending on the study. A combination of different effects seems to cause these disagreements, from the neglect of nonlocal thermodynamical equilibrium (non-LTE) corrections to the use of different atomic data \citep[see e.g.,][and references therein]{2007AJ....134.1216J,2008A&A...488..943S,2012MNRAS.422.1562S,2015MNRAS.446.3556M}. Therefore, it remains unclear whether there is agreement between stellar evolution models and observed Na and Al abundances. 

In this work, we take advantage of the Gaia-ESO Survey \citep{2012Msngr.147...25G,2013Msngr.154...47R} to study the behavior of Na and Al abundances in dwarfs and giants in the context of both stellar and Galactic chemical evolution. This work may be considered a pilot of a larger study to be conducted once the Gaia-ESO survey is completed and many more field stars and open cluster giants are observed. This paper is organized as follows. In Sect.\ \ref{sec:data} we describe the Gaia-ESO data and their analysis. In Sect.\ \ref{sec:stellar} we present a comparison of the observed Na and Al abundances in stars of open clusters with stellar evolution models. In Sect.\ \ref{sec:chemical} we discuss the comparison of the abundances with Galactic chemical evolution models. Finally, Sect.\ \ref{sec:end} summarizes our findings.


\section{Data and analysis}\label{sec:data}

\subsection{Gaia-ESO spectra and analysis}

We use Gaia-ESO Survey\footnote{\url{http://www.gaia-eso.eu}} results available in its second and third internal data releases (hereafter iDR2 and iDR3, respectively). Gaia-ESO is a public spectroscopic survey that is conducted with FLAMES \citep[Fiber Large Array Multi-Element Spectrograph,][]{2002Msngr.110....1P} at the European Southern Observatory's (ESO) Very Large Telescope (VLT) in Paranal, Chile.

The Gaia-ESO targets have different spectral types (from O type to M type) and belong to Milky Way fields and to open clusters of different ages and metallicities. Medium- ($R$ $\sim$ 20\,000) and high-resolution ($R$ $\sim$ 47\,000) spectra are obtained with the Giraffe and UVES \citep[Ultraviolet and Visual Echelle Spectrograph,][]{2000SPIE.4008..534D} spectrographs, respectively. Here, we use results of the analysis of FGK-type stars observed with UVES (the adopted Giraffe settings do not allow Na measurement). The reduction of these data is described in \citet{2014A&A...565A.113S}.

The analysis details are described in \citet{2015A&A...576A..80L}, for stars observed in young open clusters ($\leq$ 100 Myr), and in \citet{2014A&A...570A.122S}, for stars observed in the solar neighborhood and open clusters with age $>$ 100 Myr. Here, we provide only a short description of the procedure; a complete discussion is available in the publications mentioned above.

The spectrum analysis is carried out with multiple pipelines. The two main advantages of this strategy over a single pipeline approach are: 1) one single pipeline is not optimal to analyze stars in all different regions of the parameter space. With multiple pipelines, we can combine their strengths in analyzing, for example, metal-rich and metal-poor stars, dwarfs and giants, or hot and cool stars; and 2) with multiple pipelines, we can investigate the degree to which the different methods agree in each star of the sample, thus quantifying the uncertainties in a way that is not possible with the use of a single pipeline. Such a comparison of multiple pipelines gives an estimate of the precision with which the results can be obtained.

The results of each pipeline were validated using a series of calibrators (Pancino et al., in preparation), which include open and globular cluster stars and the Gaia benchmarks, a set of well-studied bright stars with fundamental atmospheric parameters \citep[][]{2014A&A...566A..98B,2014A&A...564A.133J,2015A&A...582A..49H}. For the bulk of our sample stars, analyzed as described in \citet{2014A&A...570A.122S}, the final recommended values of atmospheric parameters and abundances are weighted medians of those from the validated methods. Weights are computed with respect to the Gaia benchmarks in a procedure that ties our results to a system of reference defined by atmospheric parameters of these stars. In the Gaia-ESO catalog, each parameter is given together with an estimate of the method-to-method dispersion and the number of pipelines used for its computation.

\subsection{Sample description}\label{sec:sample}

\setcounter{table}{1}
\begin{table*}
 \caption[]{\label{tab:clusters} Properties of the open clusters for which abundances of Na and Al in giants are available.}
\centering
\begin{tabular}{lccccc}
\hline
\hline
Name & Age & [Fe/H] & $M_{\rm TO}$ & RV  & \# of \\
 &  (Gyr) & (dex) &  ($M_{\odot}$)&  (km s$^{-1}$) & giants \\
\hline
NGC 6705 & 0.316 & +0.01 $\pm$ 0.06 & 3.2 & +34.5 &  22 \\
NGC 4815 & 0.630 & $-$0.02 $\pm$ 0.04 & 2.5 & $-$30.2 & 4 \\
Berkeley 81 & 0.980 & +0.25 $\pm$ 0.08 & 2.2 & +47.6 & 13 \\
Trumpler 20 & 1.660 & +0.09 $\pm$ 0.08 & 1.8 & $-$40.5 & 40 \\
NGC 2243 & 3.5 & $-$0.44 $\pm$ 0.05 & 1.2 & +59.5 & 18 \\
Berkeley 25 & 4.5 & $-$0.27 $\pm$ 0.02 & 1.15 & +135.2 & 6 \\
\hline
\end{tabular}
\tablefoot{The mean cluster [Fe/H] values are given together with the standard deviation. The cluster RV is the mean of the giants that we consider to be cluster members. Thus, it can be slightly different from the values adopted to establish membership that are discussed in the text.}
\end{table*}

In the iDR2+iDR3 catalog, atmospheric parameters and abundances are available for 1542 FGK-type stars observed with UVES in the setup with central wavelength 580~nm. To select our sample, first we excluded stars observed in the fields of globular clusters, as their Na and Al abundances might be affected by additional processes that would introduce extra complexity in our analysis \citep[see, e.g.,][]{2012A&ARv..20...50G}. Second, we restricted the sample to stars with effective temperature ($T_{\rm eff}$) above 4000 K. In the Gaia-ESO releases used here, the results for cooler stars are less reliable because of the increased importance of line blends \citep[for stars analyzed as described in][]{2014A&A...570A.122S}. Future releases are expected to have improvements in this respect. We thus started with a sample of 1303 stars, including 1274 with Na abundances and 1246 with Al abundances. All abundances are listed in Table 1. The sample included 957 dwarfs ($\log~g$ $>$ 3.50 dex) and 346 giants ($\log~g$ $\leq$ 3.50 dex). The sample of solar neighborhood dwarfs (within $\sim$ 2 kpc of the Sun) includes mostly thin and thick disk objects, and likely few or no halo stars. We do not separate stars of the two disk components, as such a comparison is not one of our goals. The chemical differences between thin and thick disks have been studied with Gaia-ESO Giraffe data by \citet{2014A&A...567A...5R}, \citet{2014A&A...572A..33M}, and \citet{2015A&A...582A.122K}.

The full sample included stars in 16 open clusters. No distinction was made between cluster and field dwarf stars for the Galactic chemical evolution discussion (Sect.\ \ref{sec:chemical}). For the stellar evolution discussion (Sect.\ \ref{sec:stellar}), we used giants in \object{NGC 2243}, \object{NGC 4815}, \object{NGC 6705}, \object{Berkeley 25}, \object{Berkeley 81}, and \object{Trumpler 20} (Table \ref{tab:clusters}). Observations with UVES in these old and intermediate-age open clusters are focused on clump giants. The stars observed in the remaining clusters were all main-sequence or pre-main-sequence stars. We adopted the values of age and turn-off masses obtained using the PARSEC isochrones \citep{2012MNRAS.427..127B} from earlier Gaia-ESO papers for NGC 6705, NGC 4815, Berkeley 81, and Trumpler 20. For NGC 2243 and Berkeley 25, we derived ages and turn-off masses ourselves, also using PARSEC isochrones for consistency\footnote{For the fitting, we made use of the following photometric data: $VI$ data of \citet{1996A&AS..118..303K} and $BVI$ data of \citet{2005A&A...442..917C}, for NGC 2243 and Berkeley 25, respectively.}. The metallicities and mean radial velocities in Table \ref{tab:clusters} are the average of the giants that we considered to be cluster members (see discussion below). These values might be slightly different from the values published in earlier Gaia-ESO papers, which made use of the science verification iDR1. Here we used iDR2 and iDR3, new data releases made after a full reanalysis of the whole Gaia-ESO data set that, for some clusters, also included observations of additional stars.

\subsubsection{NGC 6705}

The Gaia-ESO science verification analysis of this cluster was presented in \citet{2014A&A...569A..17C}. We adopt the cluster age derived in that work using PARSEC isochrones, 0.316 Gyr, which is similar to other values in the literature, such as the 0.25 Gyr found by \citet{2013PASP..125.1412B}. A total of 49 stars of NGC 6705 were observed. Some of them were AB-type fast rotating main-sequence stars and were thus not considered in our discussion. Abundances of Na and Al were available for 24 giants with $T_{\rm eff}$ above 4000 K. We selected members adopting the mean radial velocity (RV) and dispersion determined by \citet{2014A&A...569A..17C}, i.e., +34.1 $\pm$ 1.5 km s$^{-1}$, and a three sigma criterium. We found 22 giants to be members. 

\subsubsection{NGC 4815}

The Gaia-ESO science verification analysis of NGC 4815 was presented in \citet{2014A&A...563A.117F}. Using PARSEC isochrones, they derived an age of 0.63 Gyr, which we adopt here. This agrees with the conclusion of \citet{1994A&AS..106..573C} that NGC 4815 is about the age of the Hyades. A total of 14 stars were observed in the field of the cluster. For 12 stars with $T_{\rm eff}$ $>$ 4000 K, Na and Al abundances were available. We selected members using the same RV criterium of \citet{2014A&A...563A.117F}, i.e., stars with RV = $-$29.4 $\pm$ 4.0 km s$^{-1}$ were considered to be members. Five giants satisfy this criterium, but we only used four of them and further excluded star \# 1795 as it is the most luminous and cool giant of the sample. These characteristics make the analysis of this cool giant more challenging \citep[see discussion in][]{2014A&A...563A.117F}.

\begin{figure*}
\centering
\includegraphics[height = 7cm]{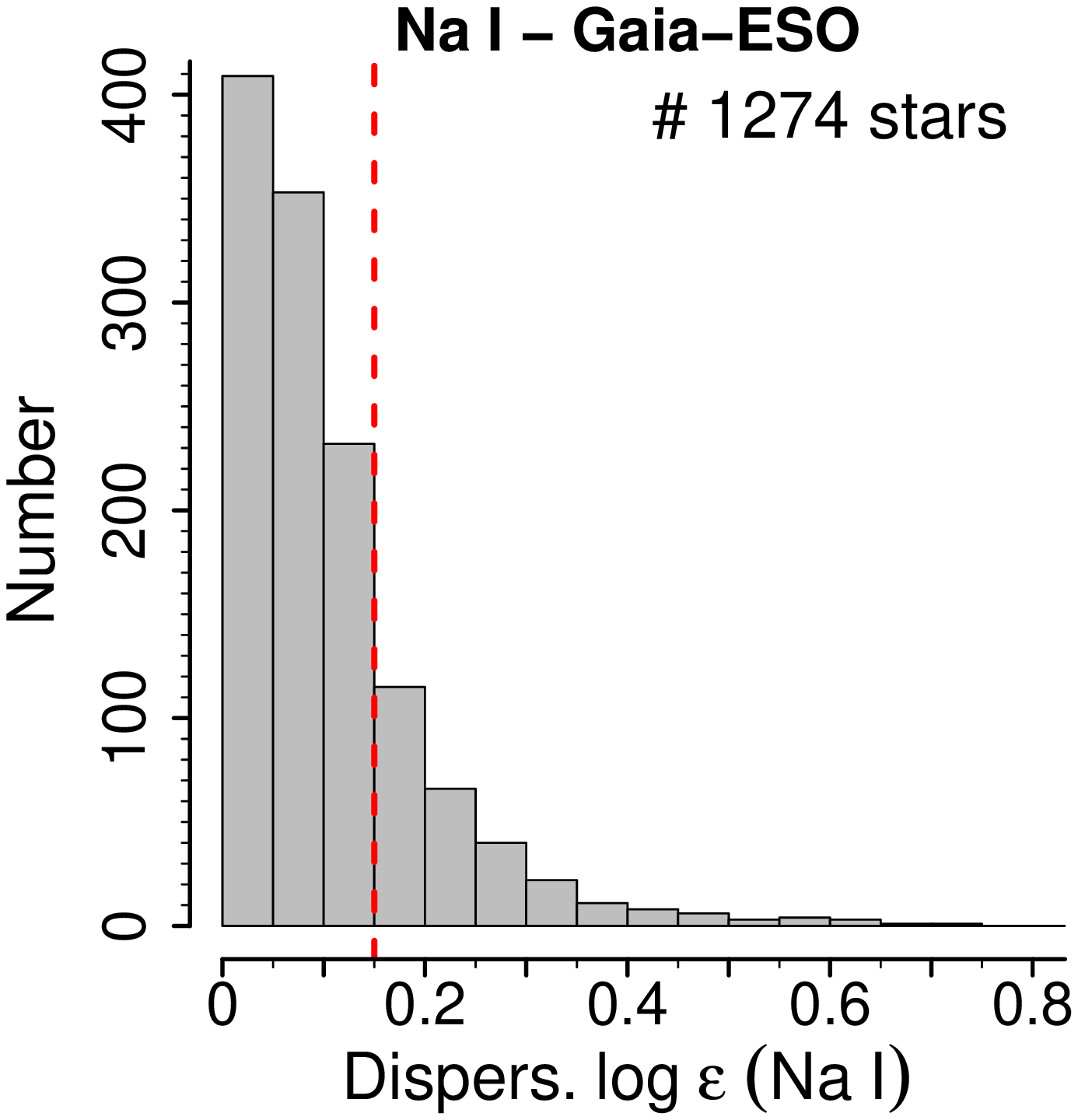}
\includegraphics[height = 7cm]{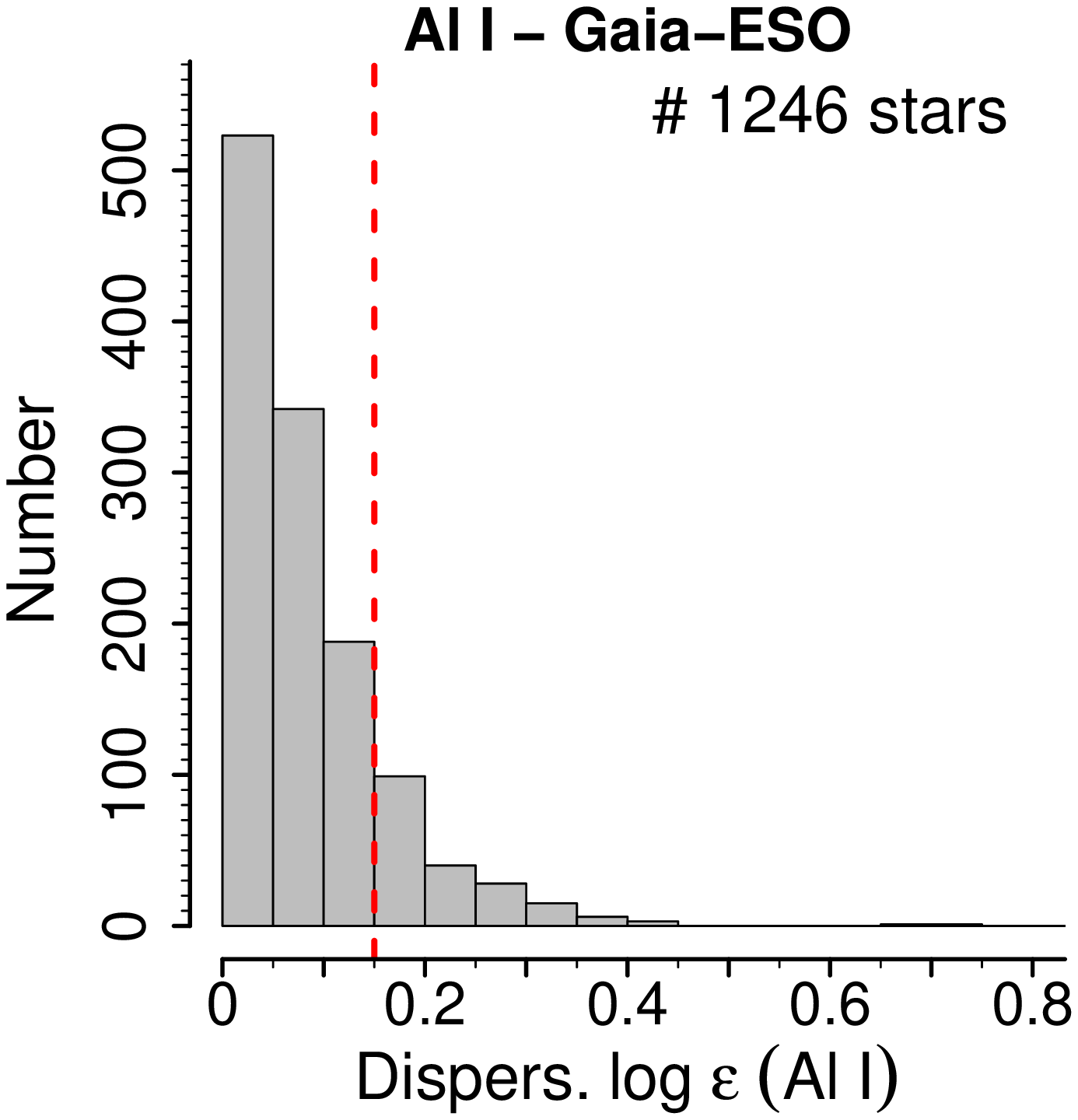}
 \caption{Precision of the Gaia-ESO Na (left panel) and Al (right panel) abundances available in the iDR2+iDR3 final catalog. The red dashed line indicates the limit of 0.15 dex used to select the best-quality abundances as described in Sect.\ \ref{sec:best}.}\label{fig:disp}%
\end{figure*}

\subsubsection{Berkeley 81}

The Gaia-ESO analysis of this cluster was presented in \citet{2015A&A...580A..85M}. The cluster age and turn-off mass found in that analysis (Table \ref{tab:abun}) are in very good agreement with those found by \citet{2014MNRAS.437.1241D}, i.e., age = 0.9 Gyr and $M_{\rm TO}$ = 2.1 $M_{\odot}$. Gaia-ESO observations were obtained for 14 giants in the field of Berkeley 81 with UVES. All stars have $T_{\rm eff}$ above 4000 K and have available abundances of both Na and Al. These 14 stars have mean RV = +47.5 $\pm$ 0.70 km s$^{-1}$, with a total range between +46.28 and +48.73 km s$^{-1}$. \citet{2014AJ....147...69H} performed a RV study of this cluster and found a mean RV = +48.1 $\pm$ 2.0, which is in excellent agreement with the value found here. These authors considered any star with RV within 5 km s$^{-1}$ of the mean value to be member. All of our 14 giants satisfied this criterium, however, the metallicity of one of them differs from the mean by more than 3$\sigma$. We considered this star to be a nonmember and excluded it from the stellar evolution discussion.
 
 \subsubsection{Trumpler 20}

The science verification results of Trumpler 20 were published in \citet{2014A&A...561A..94D}. We adopt the age that they derived using PARSEC isochrones, 1.66 Gyr, which is similar to the age of 1.5 Gyr obtained by \citet{2010NewA...15...61S}. At that time, only 13 giants had been observed and analyzed.  We presently have observations for 42 stars. A total of 41 stars have  $T_{\rm eff}$ above 4000 K and abundances of both Na and Al. We selected members using the RV criterium of \citet{2014A&A...561A..94D}, i.e., stars with RV within $-$40.4 $\pm$ 3.7 km s$^{-1}$ were considered members. A total of 40 giants were retained.

\subsubsection{NGC 2243} 
 
NGC 2243 is one of the most metal-poor open clusters known. A Gaia-ESO analysis of this cluster has not been published yet. The cluster age and turn-off mass that we derived here (Table \ref{tab:abun}) are in very good agreement with those found by \citet{2006AJ....131.1544B}, i.e., age = 4.0 Gyr and $M_{\rm TO}$ = 1.2 $M_{\odot}$. We analyzed spectra of 29 different stars: 27 observed by Gaia-ESO and two obtained from the ESO archive. Atmospheric parameters were derived for 26 of them. Based on the RVs, we considered 19 giants to be likely members (mean RV = 59.5 $\pm$ 0.8 km $^{-1}$). This value is slightly lower than the mean RV of 61.9 $\pm$ 0.8 km s$^{-1}$ found by \citet{2013A&A...552A.136F} for 82 member stars observed with the Giraffe spectrograph. We further excluded one star with a metallicity higher than that of the others ([Fe/H] = $-$0.17), leaving 18 members.
 
\subsubsection{Berkeley 25} 
 
Ten giants in the field of Berkeley 25 were analyzed. The cluster age that we derived, 4.5 Gyr, is in reasonable agreement with the age of 5 Gyr obtained by \citet{2007A&A...476..217C}. Seven were observed by Gaia-ESO, and three taken from archival data \citep[from the dataset analyzed in][]{2007A&A...476..217C}. Abundances were available for nine of them. Two of the stars have an RV that is somewhat discrepant with respect to the others; RV = 146.5 and 111.6 km s$^{-1}$ compared to a mean RV = 135.1 $\pm$ 0.8 km s$^{-1}$ (without the two). A third star seemed to have a somewhat discrepant metallicity ([Fe/H] = $-$0.41) when compared to the remaining stars (mean of $-$0.23 $\pm$ 0.06, computed without the discrepant star). This star also has the lowest $\log~g$ of the sample, therefore increased systematic errors in its parameters cannot be excluded. While we prefer not to draw strong conclusions about membership here, and defer it to a forthcoming publication, for our discussion we considered that only the remaining six giants are members of the cluster. 

\begin{table*}
 \caption[]{\label{tab:abun} Mean abundances of Na and Al (with standard deviation) in the giants of each open cluster, after the selection of the best-quality values.}
\centering
\begin{tabular}{lccccc}
\hline
\hline
Cluster & [Na/Fe] & [Na/Fe] & \# giants with & [Al/Fe]  & \# giants with  \\
 & (LTE) & (non-LTE) & good Na abun. & (LTE) & good Al abun. \\
\hline
NGC 6705 & 0.42 $\pm$ 0.11 & 0.38 $\pm$ 0.11 & 7 & 0.30 $\pm$ 0.04  & 18 \\
NGC 4815 & 0.17 $\pm$ 0.09 & 0.13 $\pm$ 0.09 & 3 & 0.06 $\pm$ 0.05 & 4 \\
Berkeley 81 & 0.27 $\pm$ 0.06 & 0.22 $\pm$ 0.06 & 8 & 0.10 $\pm$ 0.04 & 12 \\
Trumpler 20 & 0.09 $\pm$ 0.06 & 0.06 $\pm$ 0.06 & 31 & 0.02 $\pm$ 0.03 & 38 \\
NGC 2243 & 0.10 $\pm$ 0.07 & 0.10 $\pm$ 0.07 & 15 & 0.07 $\pm$ 0.05 & 17 \\
Berkeley 25 & 0.05 & 0.04  & 1 & 0.05 $\pm$ 0.01 & 4 \\
\hline
\end{tabular}
\end{table*}

\begin{figure*}
\centering
\includegraphics[height = 7cm]{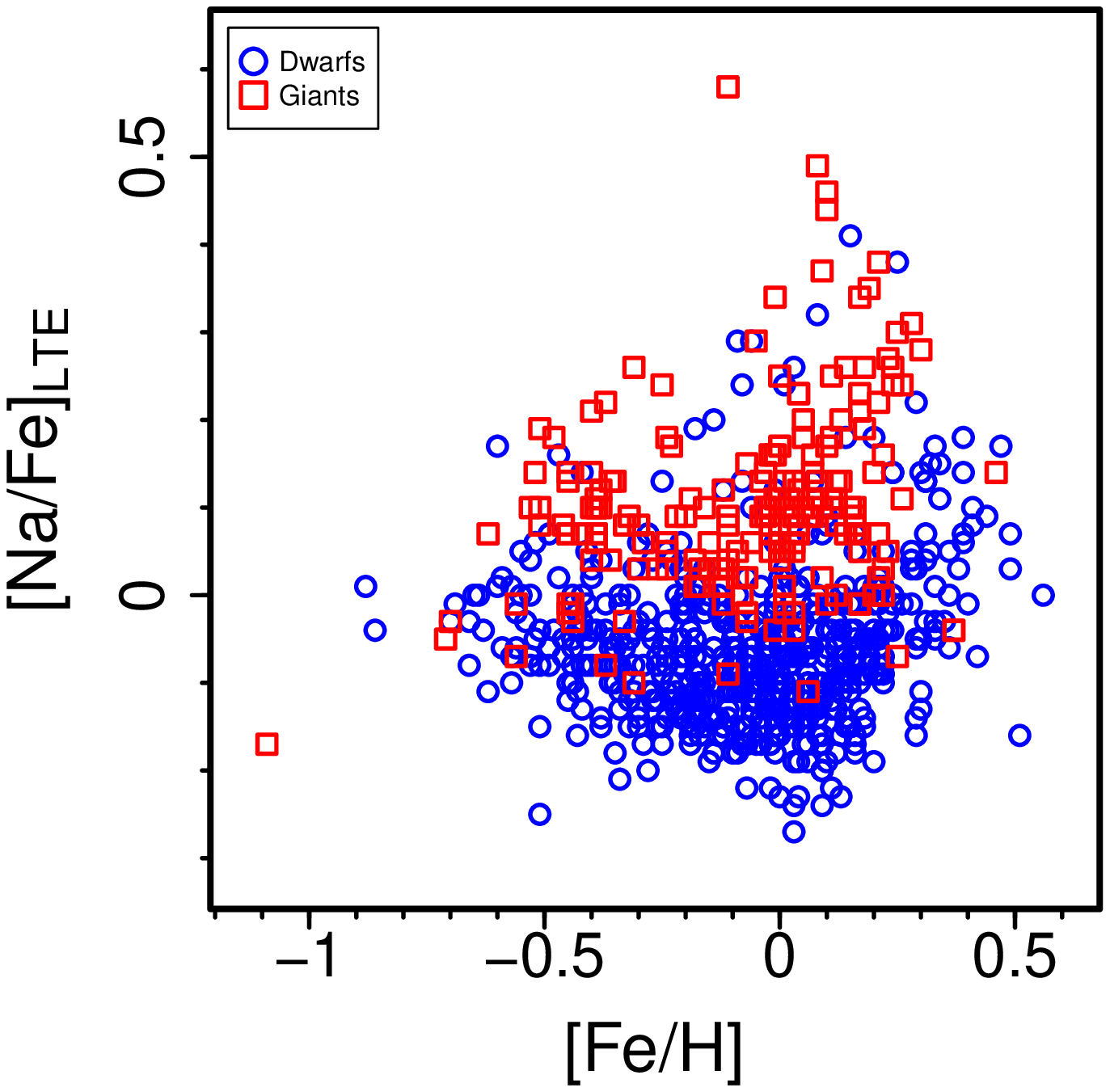}
\includegraphics[height = 7cm]{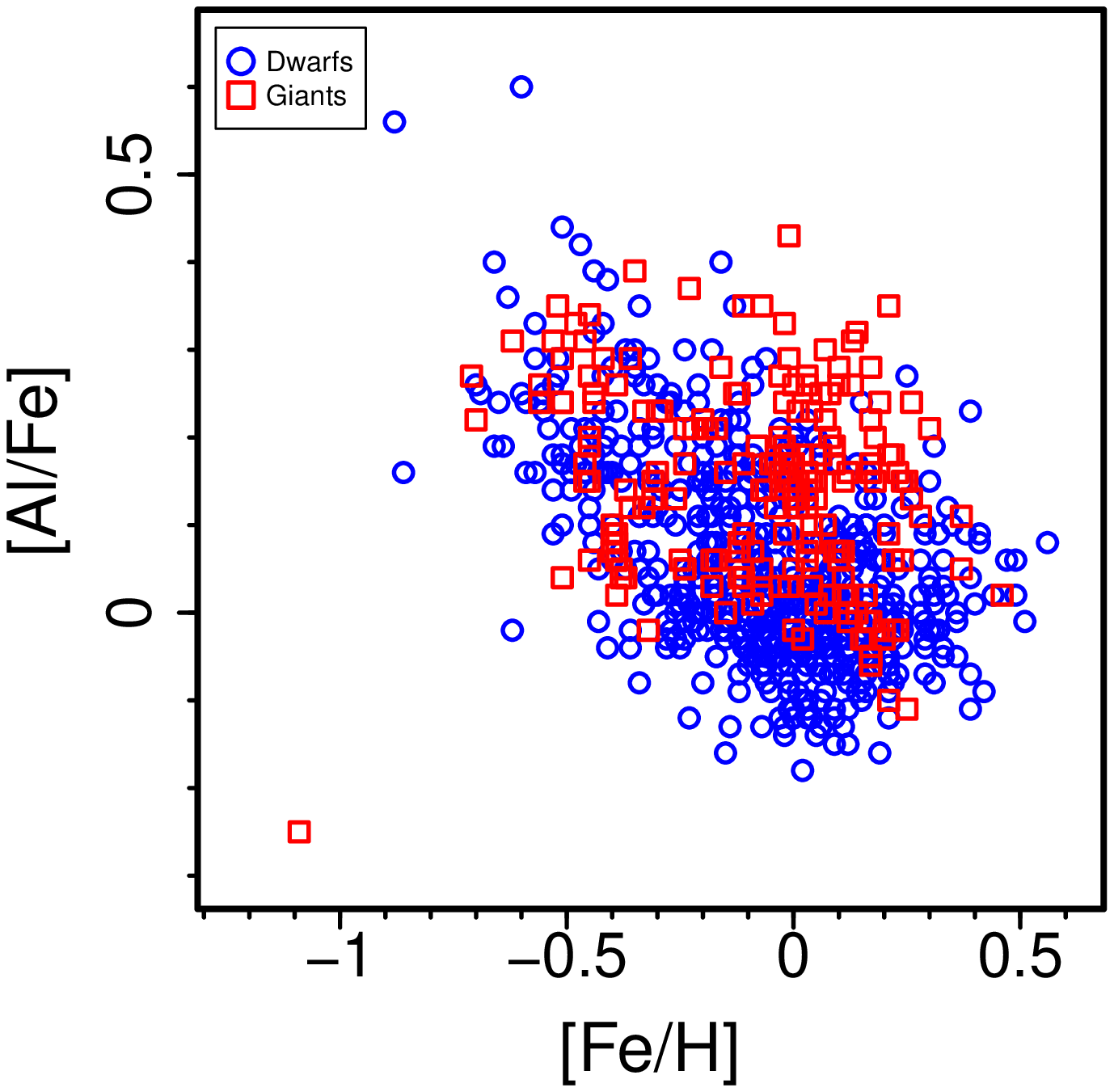}
 \caption{Sodium and aluminium abundances as a function of metallicity for the high-quality sample (Section \ref{sec:best}). Dwarfs are shown as blue circles and giants as red squares. Some remaining systematic difference between dwarfs and giants might be present, as discussed in Section \ref{sec:sys}.}\label{fig:abunfeh}%
\end{figure*}

\subsection{Selecting the best-quality abundances}
\label{sec:best}

A total of up to six Na lines ($\lambda$ 4982.814, 5153.402, 5682.633, 5688.205, 6154.226, and 6160.747 \AA) and up to three Al lines ($\lambda$ 5557.063, 6696.023, 6698.673 \AA) were used to compute the Gaia-ESO abundances. The atomic data were part of the version 4 of the Gaia-ESO line list, details of which will be published elsewhere \citep[see a discussion in][]{2015PhyS...90e4010H}. 

The recommended Gaia-ESO abundances are given in the $\log \epsilon$ format\footnote{$\log \epsilon$(X) = $\log$ [N(X)/N(H)] + 12, i.e., a logarithmic abundance by number on a scale where the number of hydrogen atoms is 10$^{12}$.}. For each star, a weighted median of the multiple pipeline results, on a line-by-line basis, was computed. The final abundance was then the median of all the line abundances. The median absolute deviation is used as a measurement of the dispersion and can be understood as the precision of the results \citep[see][]{2014A&A...570A.122S}. In Fig.\ \ref{fig:disp} we show the histograms of these dispersions for the Na and Al abundances in our sample.

There is an extended tail in the dispersion distribution reaching values above 0.40 dex for both Na and Al. The third quartile of the dispersion distribution is 0.14 dex for Na and 0.12 dex for Al. We thus decided to use only abundances with dispersion in Na or Al $\leq$ 0.15 dex to remove the more uncertain results.

In addition, because of how the recommended parameters and abundances are obtained in Gaia-ESO, the number of pipelines on which the results are based is also important. In Section 7.6 of \citet{2014A&A...570A.122S}, it is discussed how the accuracy of the recommended atmospheric parameters changes with the use of results from different numbers of pipelines. Similar arguments apply to the accuracy of the abundances. 

Results based on fewer pipelines have an increased potential to be more uncertain and thus increase the scatter of the values in the sample. Robust recommended abundances are those based on many determinations, as this guarantees that the distribution of pipeline results (affected by random uncertainties) is well sampled and outlier results are properly identified. In \citet{2014A&A...570A.122S}, it was shown that selecting recommended values based on at least five pipelines would guarantee that the majority of the selected results was close to the best possible values. 

Here, we decided to use abundances based on determinations from at least four different pipelines. This was a compromise needed to avoid losing too many stars from the sample, which have values coming from only four different pipelines. The effect of stopping at four and not five pipelines will be an increase in the scatter of our abundances. We do not expect the choice to introduce any bias in the results. The abundances in Table 1 are given together with the abundance dispersion and number of pipelines on which they are based.

Of the 13 different pipelines available during the analysis, up to eight provided abundances of Na and Al, although not for all stars. Internally to Gaia-ESO these eight pipelines are known as Bologna, CAUP, Concepcion, EPINARBO, LUMBA, Paris-Heidelberg, UCM, and Vilnius, and are described in Appendix A of \citet{2014A&A...570A.122S}.

The restrictions above reduce the sample to 908 stars with Na abundances ($\sim$ 71\% of the original 1274 stars): 237 giants and 631 dwarfs. The sample with Al abundances is reduced to 941 stars ($\sim$ 75\% of the original 1246 stars): 252 giants and 689 dwarfs. These selected abundances are shown as a function of metallicity in Fig. \ref{fig:abunfeh}. Table \ref{tab:abun} lists the mean abundances of Na and Al, and associated standard deviations, for the selected giants in each open cluster.


\subsection{Systematic effects on the abundances}
\label{sec:sys}

\begin{figure*}
\centering
\includegraphics[height = 6cm]{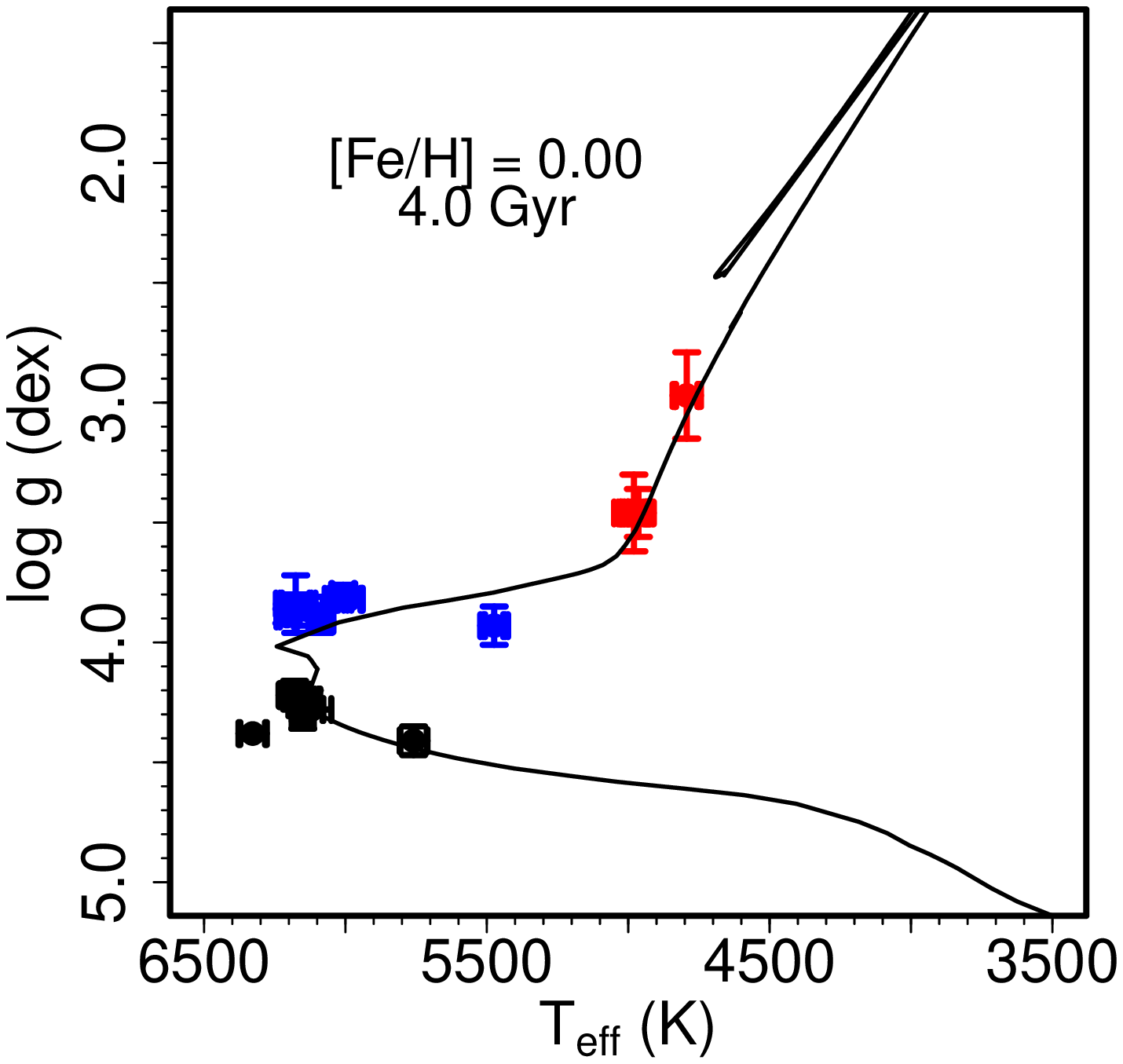}
\includegraphics[height = 6cm]{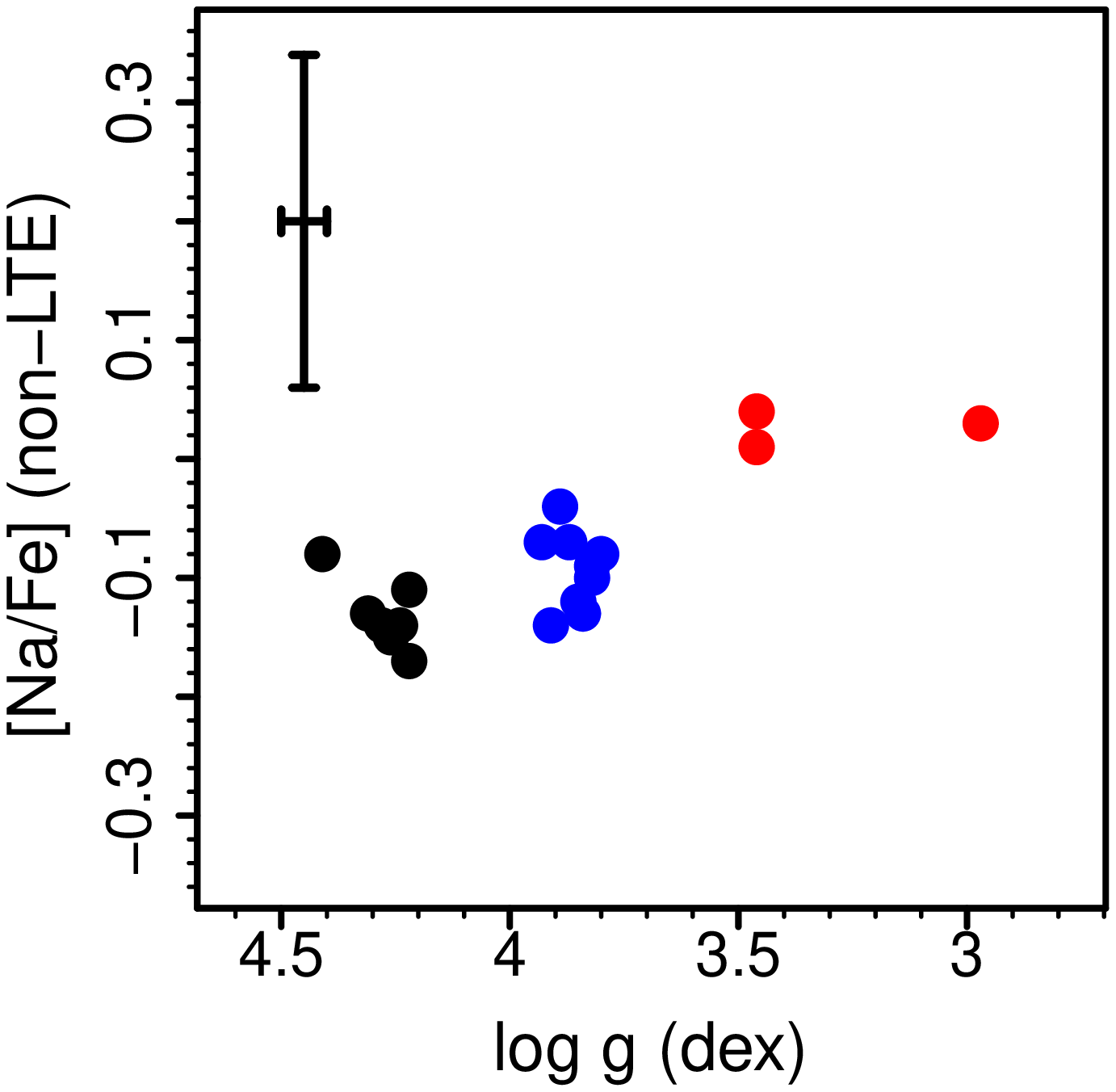}
\includegraphics[height = 6cm]{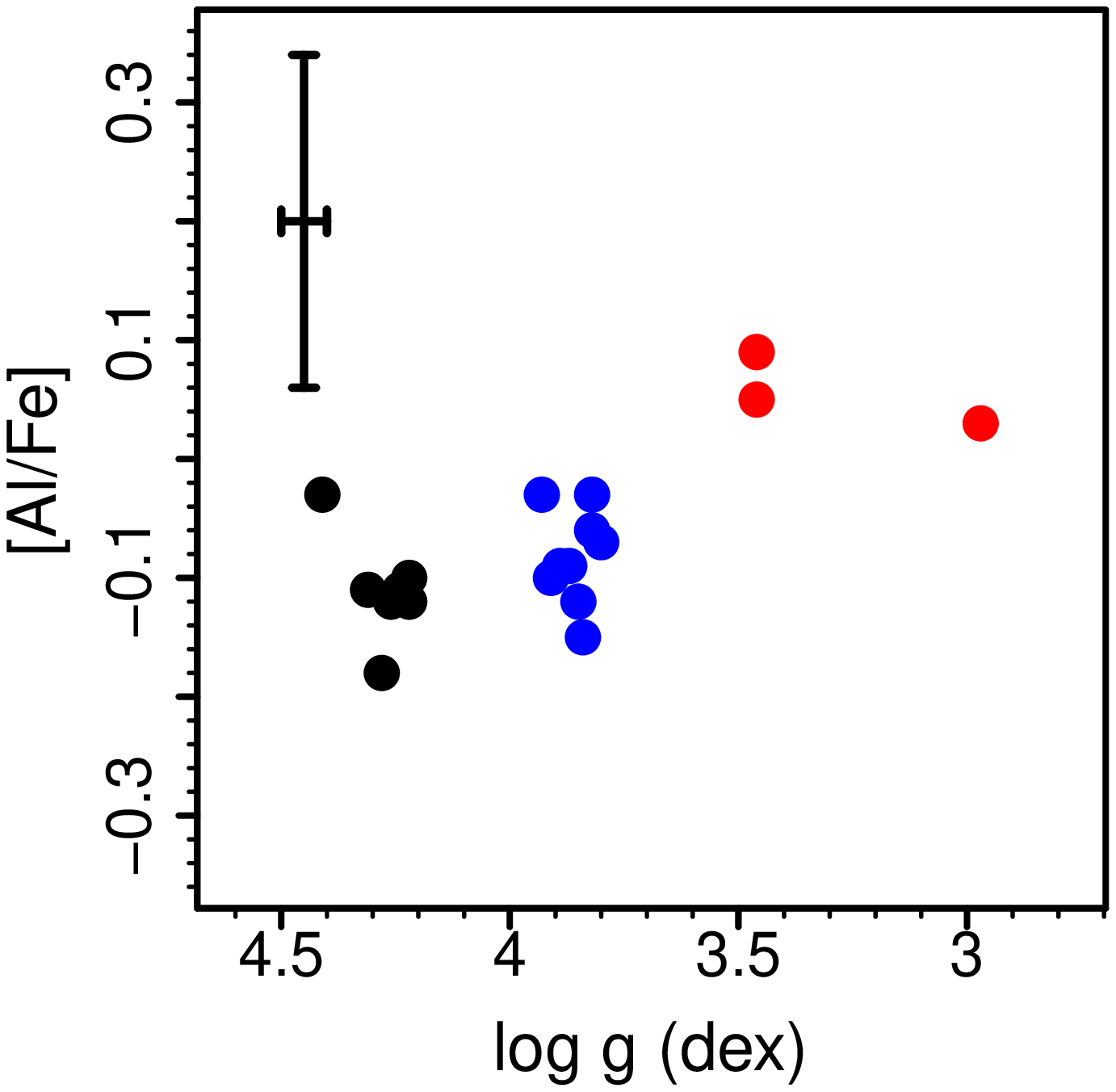}
 \caption{\emph{Left panel:} sample stars from M 67 in the $T_{\rm eff}$-$\log~g$ plane. \emph{Middle panel:} [Na/Fe] ratio of each star in M 67 as a function of its surface gravity. \emph{Right panel:} [Al/Fe] ratio of each star in M 67 as a function of its surface gravity. Stars are color-coded according to the surface gravity: red for giants with $\log~g \leq$ 3.5, blue for turn-off and subgiant stars with 3.5 $< \log~g \leq$ 4.0, and black for main-sequence stars with $\log~g >$ 4.0. A typical error bar ($\pm$ 0.14 dex for [Elem./Fe] and $\pm$ 0.05 dex for $\log~g$) is shown in the upper left part of the middle and right panels.}\label{fig:m67}%
\end{figure*}

The comparison of the abundances in dwarfs and giants, shown in Fig.\ \ref{fig:abunfeh}, suggests the possibility of systematic differences. Such differences can appear, for example, because of unidentified line blends that are stronger in a certain type of star.

\begin{figure*}
\centering
\includegraphics[height = 6cm]{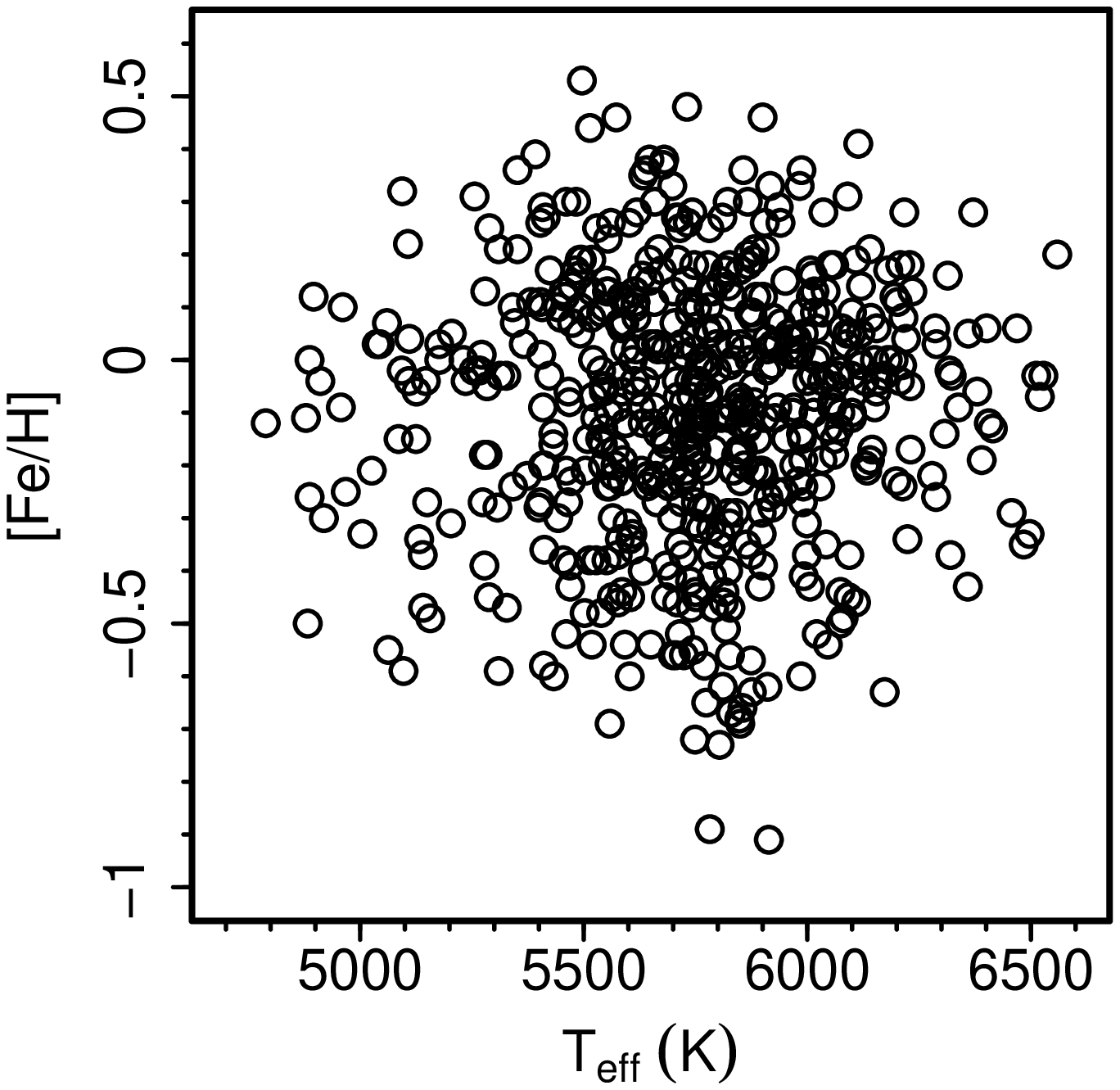}
\includegraphics[height = 6cm]{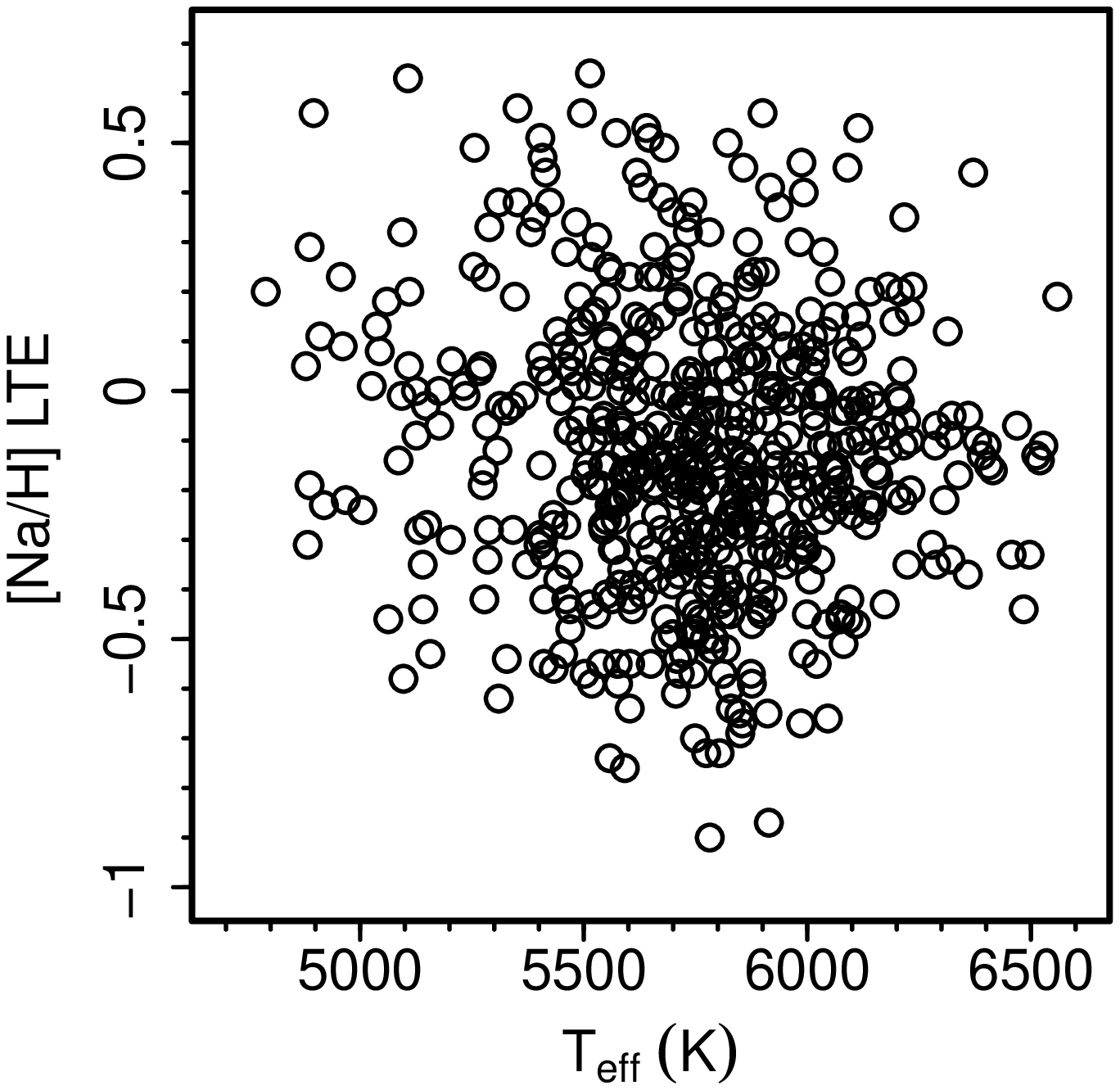}
\includegraphics[height = 6cm]{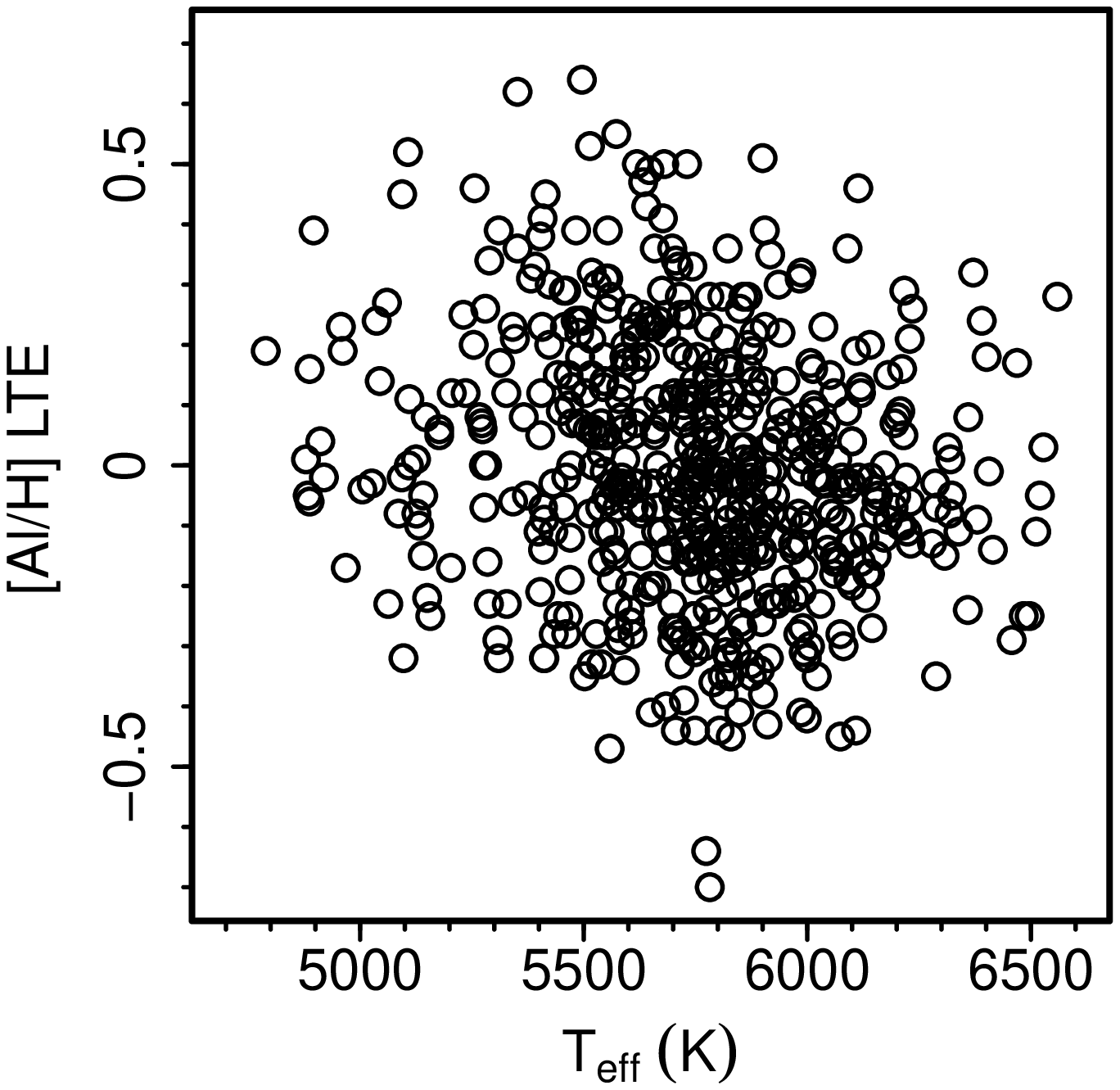}
\includegraphics[height = 6cm]{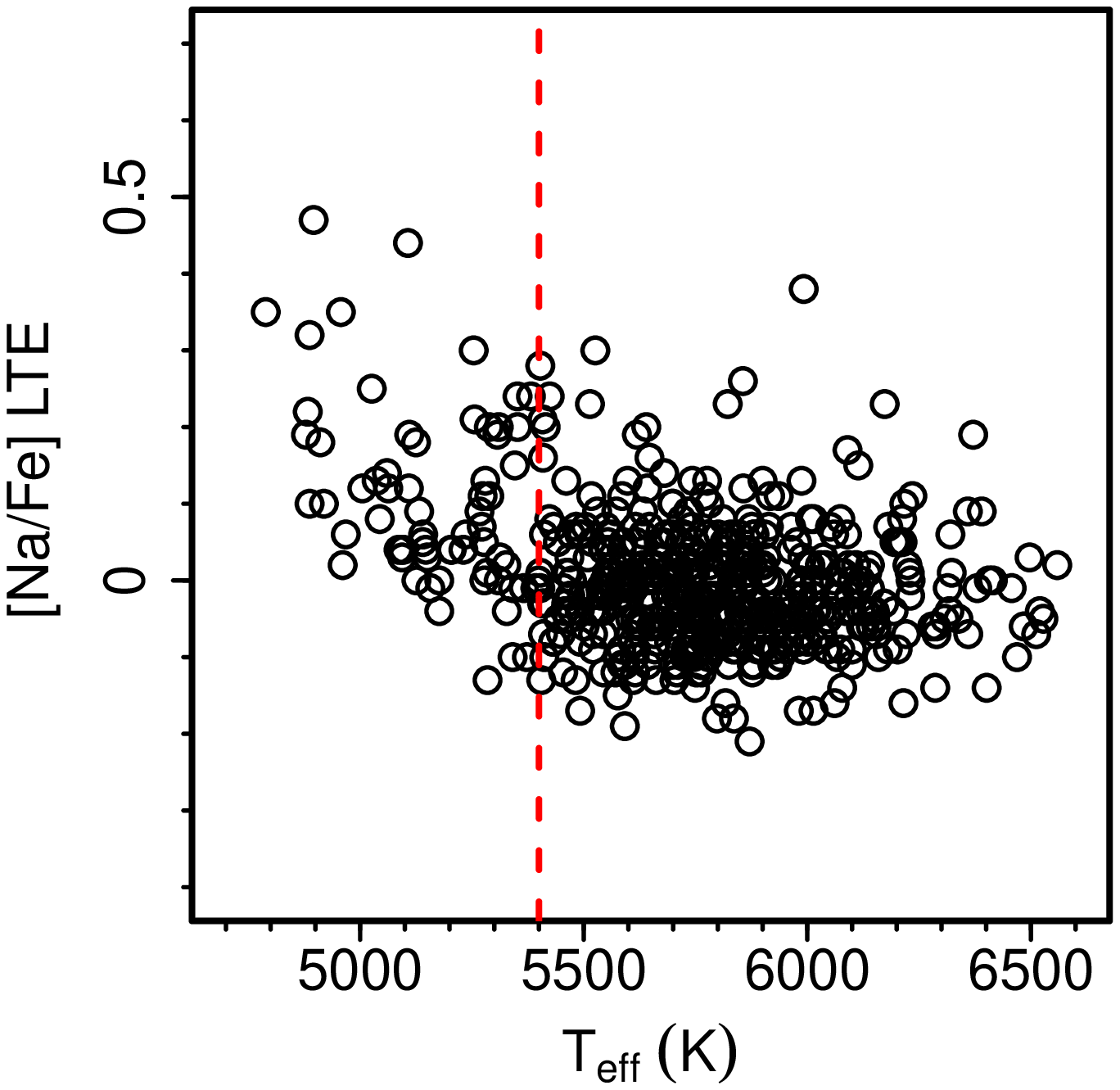}
\includegraphics[height = 6cm]{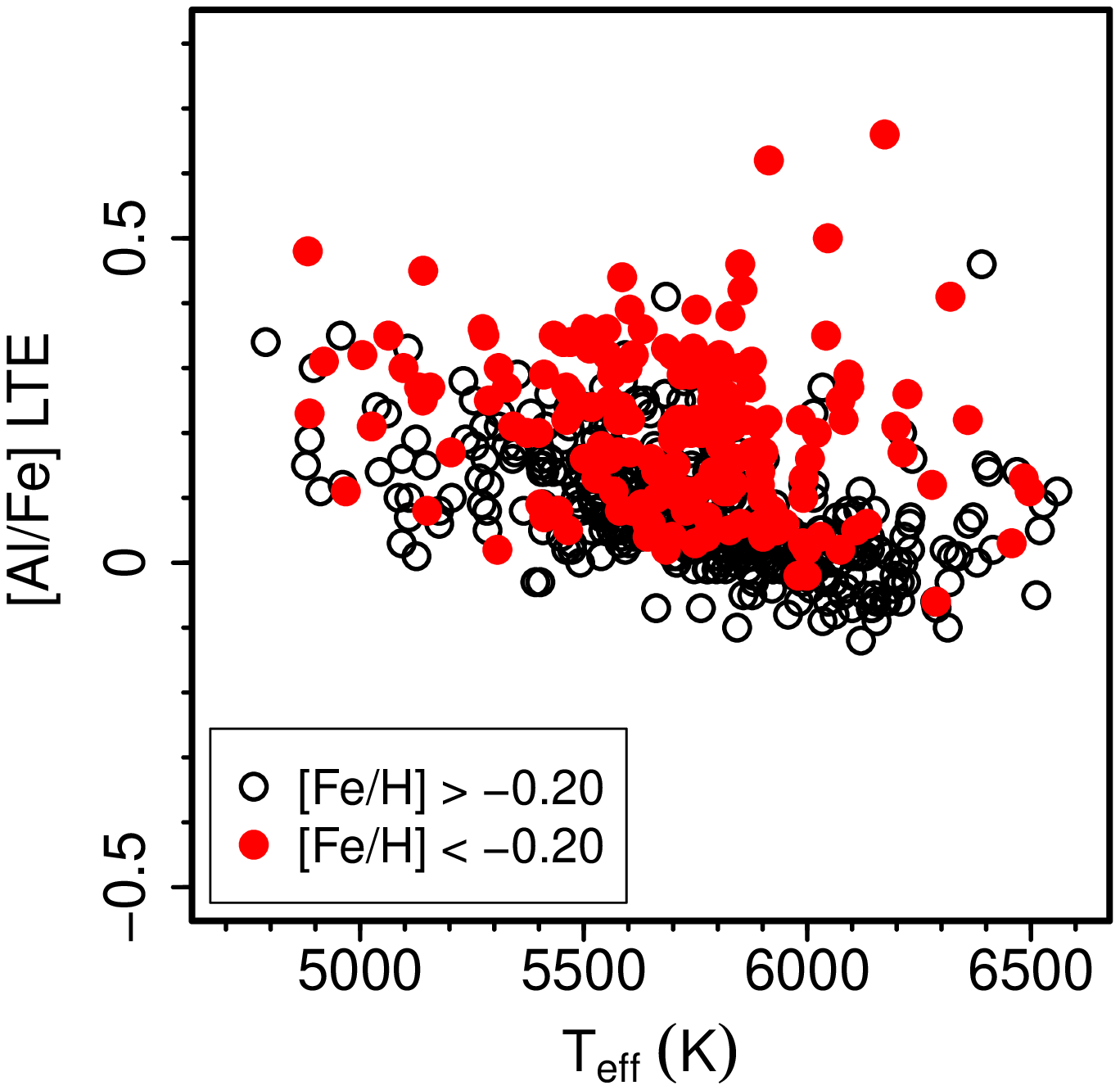}
\includegraphics[height = 6cm]{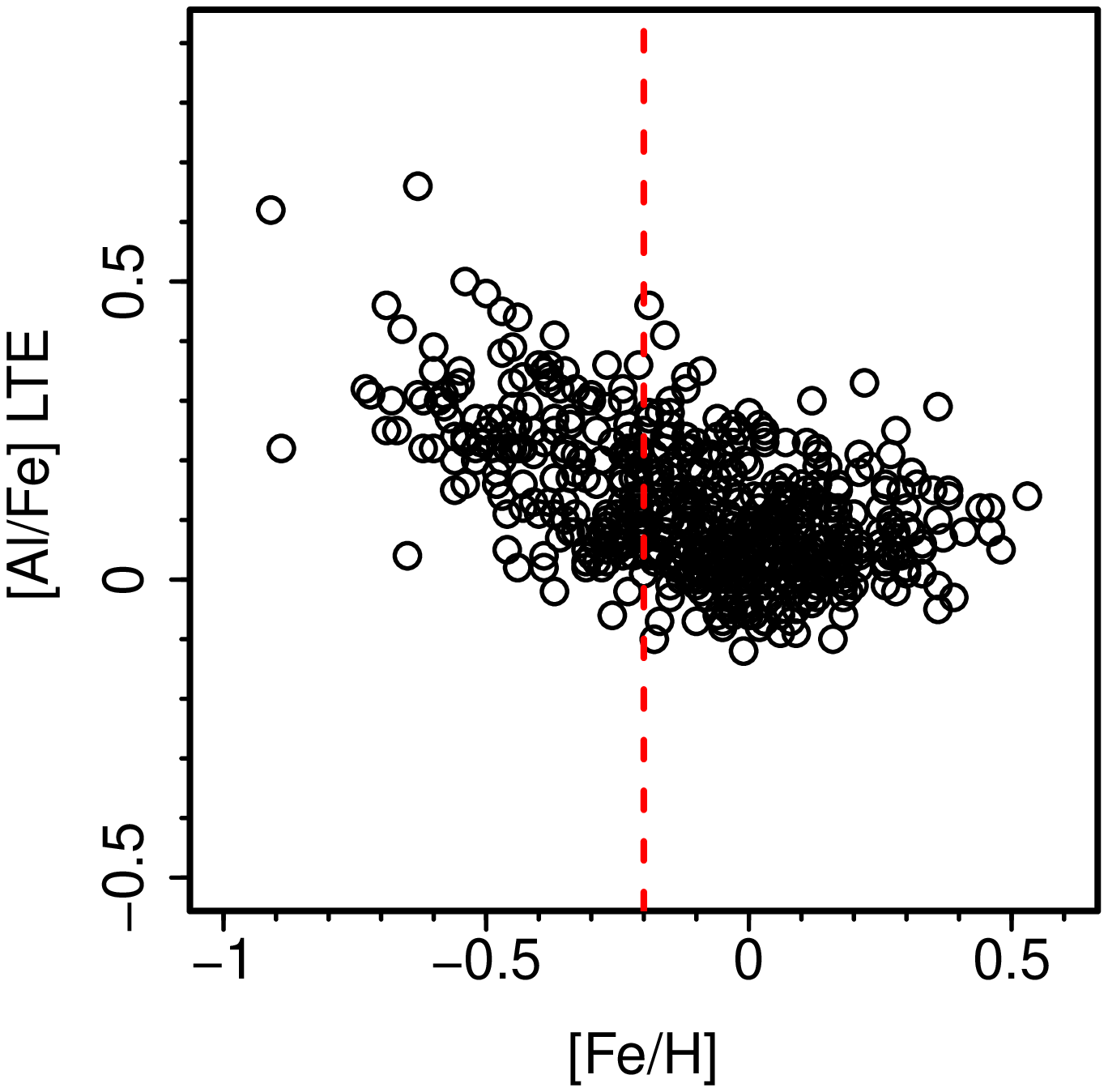}
 \caption{Trends between abundances and atmospheric parameters in the dwarfs of the sample. \emph{Upper left:} no apparent trend between [Fe/H] and $T_{\rm eff}$ (correlation coefficient $\rho$ = 0.03, with p value = 0.49). \emph{Upper middle:} weak trend between [Na/H] and $T_{\rm eff}$ ($\rho$ = $-$0.09, with p value = 0.03). \emph{Upper right:} weak trend between [Al/H] and $T_{\rm eff}$ ($\rho$ = $-$0.17, with p value close to zero). \emph{Lower left:} a moderate trend ($\rho$ = $-$0.34, with p value close to zero) between [Na/Fe] and $T_{\rm eff}$ appears for stars cooler than 5400 K (red dashed line). \emph{Lower middle:} A moderate trend between [Al/Fe] and $T_{\rm eff}$ ($\rho$ = $-$0.37, with p value close to zero). In red circles, we show the stars with [Fe/H] $< -0.20$ selected to understand whether the correlation with $T_{\rm eff}$ would affect the chemical evolution discussion. \emph{Lower right:} the trend between [Al/Fe] and [Fe/H] thought to appear from the Galactic chemical evolution. We selected the stars with [Fe/H] $< -0.20$ (left of the red dashed line) to test whether they are the stars mainly affected by the correlation of [Al/Fe] with $T_{\rm eff}$.}\label{fig:sys}%
\end{figure*}

Weak to moderate trends are indeed present between the LTE [Na/Fe] and [Al/Fe] ratios with both $T_{\rm eff}$ and $\log~g$. This leads to abundance differences between giants and dwarfs that are clearly seen, for example, in stars belonging to M 67 (Fig.\ \ref{fig:m67}). In this cluster, we have a good coverage of the evolutionary track from the main sequence to the red giant branch (RGB). Even though the giants of M 67 are located before the end of the first dredge-up, they show higher [Na/Fe] and [Al/Fe] ratios in comparison with the dwarfs. The [Fe/H] ratio does not show such a trend. This supports that some Na and Al differences between dwarfs and giants are likely caused by systematic problems in the analysis.
 
Because of this, we prefer to be cautious and avoid any discussion comparing the abundances of dwarfs to those of giants. The possibility of systematic differences between dwarfs and giants is being investigated and, if present, will be corrected in new Gaia-ESO data releases.

Within the cluster giants only, there is no correlation between [Al/H] and $T_{\rm eff}$ or $\log~g$, either between [Na/H] and $\log~g$. The metallicity seems to show a weak correlation with $T_{\rm eff}$ and $\log~g$ for stars in Trumpler 20 only. A weak correlation between [Na/H] and $T_{\rm eff}$ is also apparent in Trumpler 20, and suggested by one star in NGC 6705 (but this could eventually be a nonmember outlier). In any case, these weak correlations do not bias the stellar evolution discussion, in which we make use of average abundance ratios per cluster.

Within the dwarfs, there is no trend of [Fe/H] with $T_{\rm eff}$, but weak trends are suggested between [Fe/H] and $\log~g$, and between [Na/H] and [Al/H] and both $T_{\rm eff}$ and $\log~g$ (top row of Fig.\ \ref{fig:sys}). These weak trends are mostly imperceptible by eye, as there is a large scatter at each value of the atmospheric parameters. For [Al/Fe] and [Na/Fe], no trend is apparent with $\log~g$. However, moderate trends appear when looking at both [Na/Fe] and [Al/Fe] as a function of $T_{\rm eff}$ (bottom row of Fig.\ \ref{fig:sys}). 

For Na, the only effect of the stars with $T_{\rm eff} < 5400$ K in the [Na/Fe] vs.\ [Fe/H] plot is to increase the scatter. No systematic bias is introduced. Nevertheless, we decided to exclude such cool stars from the discussion of the chemical evolution of Na, as it is simple enough to include a temperature cut in the sample. For Al, however, the trend of [Al/Fe] with $T_{\rm eff}$ is not restricted to a given temperature range. However, as can be seen in the bottom row of Fig.\ \ref{fig:sys}, the rise of [Al/Fe] at low metallicities is not caused by the systematic trend with $T_{\rm eff}$. It seems again that this trend with temperature only affects the scatter of the points, and does not introduce further systematic effects in the interpretation of the chemical evolution of Al. We therefore do not include additional restrictions in the sample of dwarfs with Al abundances.

\begin{table*}
 \caption[]{\label{tab:sun} Results of multiple analyses of solar spectra in the Gaia-ESO Survey for the iDR2 and iDR3 cycles.}
\centering
\begin{tabular}{lccccccc}
\hline
\hline
Spectrograph & Cycle & $T_{\rm eff}$ & $\log~g$ & [Fe/H] & $\xi$ & $\log \epsilon$(Na) & $\log \epsilon$(Al)  \\
                       &           &  (K)      & (dex) & (dex) & (km s$^{-1}$) & (dex) & (dex) \\
\hline
FLAMES/UVES & iDR2 & 5826 $\pm$ 40 & 4.50 $\pm$ 0.05 & $-$0.03 $\pm$ 0.12 & 1.05 $\pm$ 0.00 & 6.31 $\pm$ 0.05 & 6.44 $\pm$ 0.01 \\
FLAMES/UVES & iDR3 & 5797 $\pm$ 85 & 4.45 $\pm$ 0.11 & 0.03 $\pm$  0.03 & 0.70 $\pm$ 0.25 & 6.27 $\pm$ 0.07 & 6.43 $\pm$ 0.09 \\
NARVAL & iDR2 & 5810 $\pm$ 17 & 4.50 $\pm$  0.08 & 0.00 $\pm$ 0.04 & 1.06 $\pm$ 0.08 & 6.29 $\pm$ 0.02 & 6.46 $\pm$ 0.01 \\
NARVAL & iDR3 & 5785 $\pm$ 40 & 4.44 $\pm$ 0.14 & 0.03 $\pm$ 0.12 & 0.94 $\pm$ 0.20 & 6.30 $\pm$ 0.13 & 6.43 $\pm$ 0.04 \\
UVES standalone & iDR2 & 5777 $\pm$ 31 & 4.43 $\pm$ 0.13 & 0.00 $\pm$ 0.06 & 1.04 $\pm$ 0.16 & 6.28 $\pm$ 0.02 & 6.45 $\pm$ 0.03 \\
UVES standalone & iDR3 & 5774 $\pm$ 25 & 4.43 $\pm$ 0.10 & 0.04 $\pm$ 0.04 & 0.95 $\pm$ 0.17 & 6.33 $\pm$ 0.07 & 6.44 $\pm$ 0.05 \\
\hline
Average & -- & 5795 $\pm$ 20 & 4.46 $\pm$ 0.03 & 0.01 $\pm$ 0.03 & 0.96 $\pm$ 0.14 & 6.30 $\pm$ 0.02 & 6.44 $\pm$ 0.01 \\
\hline
\end{tabular}
\tablefoot{The metallicities, [Fe/H], are given with respect to $\log \epsilon$(Fe) = 7.45 from \citet{2007SSRv..130..105G}.}
\end{table*}

\subsection{The solar abundances}
\label{sec:sun}

As solar reference abundances we used the results obtained from the analysis of the FLAMES/UVES solar spectrum\footnote{\url{http://www.eso.org/observing/dfo/quality/GIRAFFE/pipeline/solar.html}} in Gaia-ESO iDR2, as presented in \citet{2014A&A...570A.122S}, and listed in the first line of Table \ref{tab:sun}.

However, Gaia-ESO also analyzed a solar NARVAL\footnote{NARVAL is a spectropolarimeter on the 2m Telescope Bernard Lyot (TBL) atop Pic du Midi \citep{2003EAS.....9..105A}.} spectrum and the UVES (obtained in standalone mode) spectrum from the Gaia benchmark stars library \citep{2014A&A...566A..98B}. The three spectra were analyzed once during iDR2 and again for iDR3. With these multiple analyses (Table \ref{tab:sun}), we were able to investigate the uncertainties on our solar reference abundances. 

This comparison reveals a variation of up to 0.07 dex in [Fe/H], of up to 0.06 dex in $\log \epsilon$(Na), and of up to 0.03 dex in $\log \epsilon$(Al). These differences reflect the use of different pipelines to define the recommended parameters of the Sun in each Gaia-ESO internal release, which is an effect similar to what was discussed in Section \ref{sec:best}. We remark that the Sun was analyzed as any other star in our sample. The Sun is used as one of the benchmark stars, but no special weight is given to its analysis with respect to the other benchmarks. Thus, by itself the Sun does not define our system of parameters and abundances, but is one of the stars defining that system. The differences in the solar parameters and abundances as listed in Table \ref{tab:sun} do not reflect changes in the scales as much as they quantify uncertainties inherent in our method of defining the recommended results. Nevertheless, a solar analysis can be used as a special reference when we need to list abundances in the [Element/Fe] format. We prefer this approach, over adopting reference solar abundances from literature compilations, as our own solar analyses reflect better shortcomings such as lack of non-LTE corrections.

The important observation from Table \ref{tab:sun} is that most values (parameters and abundances) agree with each other, within their uncertainties. However, the variation in the solar abundances has an impact on the abundance ratios of the sample. Depending on which solar analysis is used as reference, there can be a maximum change of up to 0.10 dex in [Na/Fe] and up to 0.07 dex in [Al/Fe]. This is an intrinsic uncertainty of the zero point of our abundance scale and is important when comparing the element ratios to the predictions of the stellar evolution models. Nevertheless, this has no influence on the relative comparison between stars of the sample, as a zero point change would affect all stars in the same way. Moreover, we normalize the chemical evolution model predictions to our adopted solar abundances, therefore the comparison with the observations in this case is also not affected.

\subsection{Non-LTE corrections}
\label{sec:NLTE}

The Na abundances were corrected for non-LTE effects using the grids of \citet{2011A&A...528A.103L}. The corrections were derived on a line-by-line basis, using the atmospheric parameters and LTE Na abundance of each star as input. Two of the lines used in Gaia-ESO, $\lambda$ 4982.814 and 5153.402 \AA, were not part of the original grid. Nevertheless, they were part of the model atom of that work and thus the corrections for them could be computed.

The average corrections for all stars (giants and dwarfs) are always negative and range from $-$0.06 down to $-$0.18 dex. For a few stars, the non-LTE correction was actually extrapolated. This was the case for stars with $\xi <$ 1.00 km s$^{-1}$ (119 stars) and for two stars with [Fe/H] $>$ +0.50 dex, as these values are also outside the original grid. For a few other stars (24), it was not possible to compute non-LTE corrections because their LTE Na abundances had values outside the \citet{2011A&A...528A.103L} grid. For our reference Sun, the non-LTE correction is of $-$0.08 dex and the non-LTE abundance of Na is thus $\log \epsilon$(Na) = 6.23. In Table 1, we give the non-LTE [Na/Fe] ratio for those stars with good quality abundances (as described in Sect.\ \ref{sec:best}).

We did not correct the Al abundances for non-LTE effects. Although non-LTE abundances of Al have been computed in the literature \citep[e.g.,][]{1997A&A...325.1088B,2004A&A...413.1045G,2008A&A...481..481A}, no comprehensive grid of corrections is currently available for the metallicity range of our sample. 

Instead, we estimated non-LTE corrections for Al from new preliminary computations performed by one of us (T. Nordlander). The calculations were carried out for $\lambda$ 5557 and 6696--6698\,\AA\ lines and for stellar parameters and abundances representative of the giants in the open clusters, for the Sun and for two additional sets of dwarf-like parameters. The non-LTE model, which will be described in an upcoming paper (Nordlander et al., in prep.), adopts realistic hydrogen collisional rates \citep{2013A&A...560A..60B,1991JPhB...24L.127K} as well as newer electron collisional rates. 

The average corrections for the giants seem to be approximately $-$0.05 dex. We are not aware of other published non-LTE corrections for Al in solar-metallicity giants, and thus cannot compare with previous results. For the Sun and the two dwarfs, the average corrections are small, about $-$0.01/$-$0.02 dex. This agrees with the results of \citet{1996A&A...307..961B}, who found that abundances derived from the lines at 6696 and 6698 \AA\ agree in LTE and non-LTE to within 0.01 dex in the Sun. The different corrections for dwarfs and giants are not sufficient to explain the difference in [Al/Fe] between the types of stars that were discussed in Section \ref{sec:sys}.

\subsection{Ages for field dwarfs}
\label{sec:ages}

Ages and masses were computed for field dwarf stars following the procedure described in \citet{2014A&A...565A..89B}. This is accomplished with the Bellaterra Stellar Parameter Pipeline \citep{2013MNRAS.429.3645S}, which adopts a grid of stellar evolutionary tracks computed with the GARSTEC code \citep[GARching STellar Evolution Code,][]{2008Ap&SS.316...99W}. As in \citet{2014A&A...565A..89B}, we only use ages and masses obtained for stars with $\log~g$ $>$ 3.5, as the models are degenerate outside this regime, and for which the fractional age error is $<$ 30\%. The age accuracy is of course limited by the accuracy of the atmospheric parameters used in its computation and by the accuracy of the stellar models. The age values are available for 381 dwarfs in our sample. We note the use of different stellar models to compute the ages of open clusters and field stars, which likely results in two different scales. However, these two sets of ages are not discussed together. Moreover, the errors in age among the field stars are likely larger than any systematic between the two stellar models.

\section{Stellar evolution with open cluster stars}\label{sec:stellar}

\begin{figure*}
\centering
\includegraphics[height = 8cm]{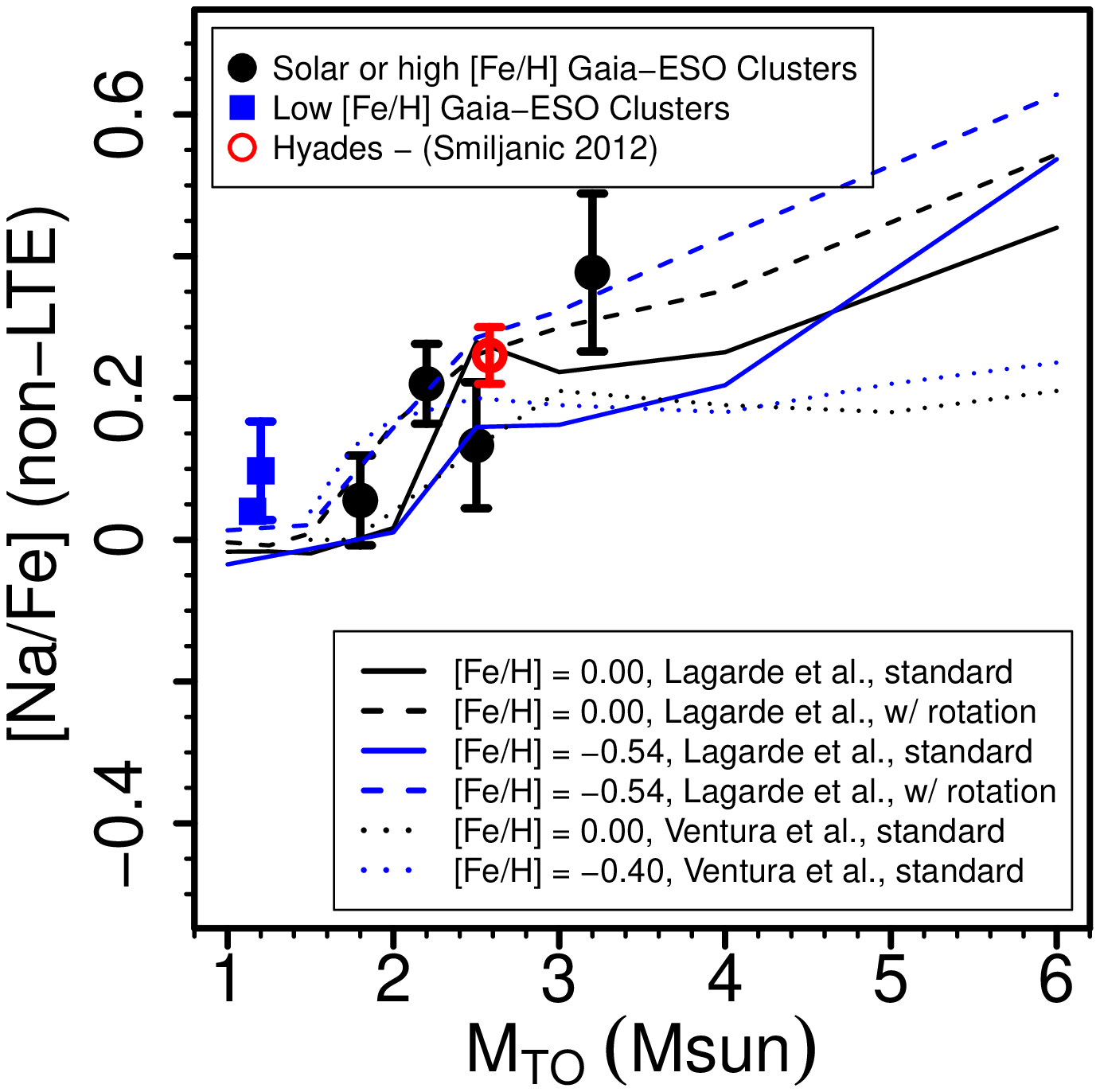}
\includegraphics[height = 8cm]{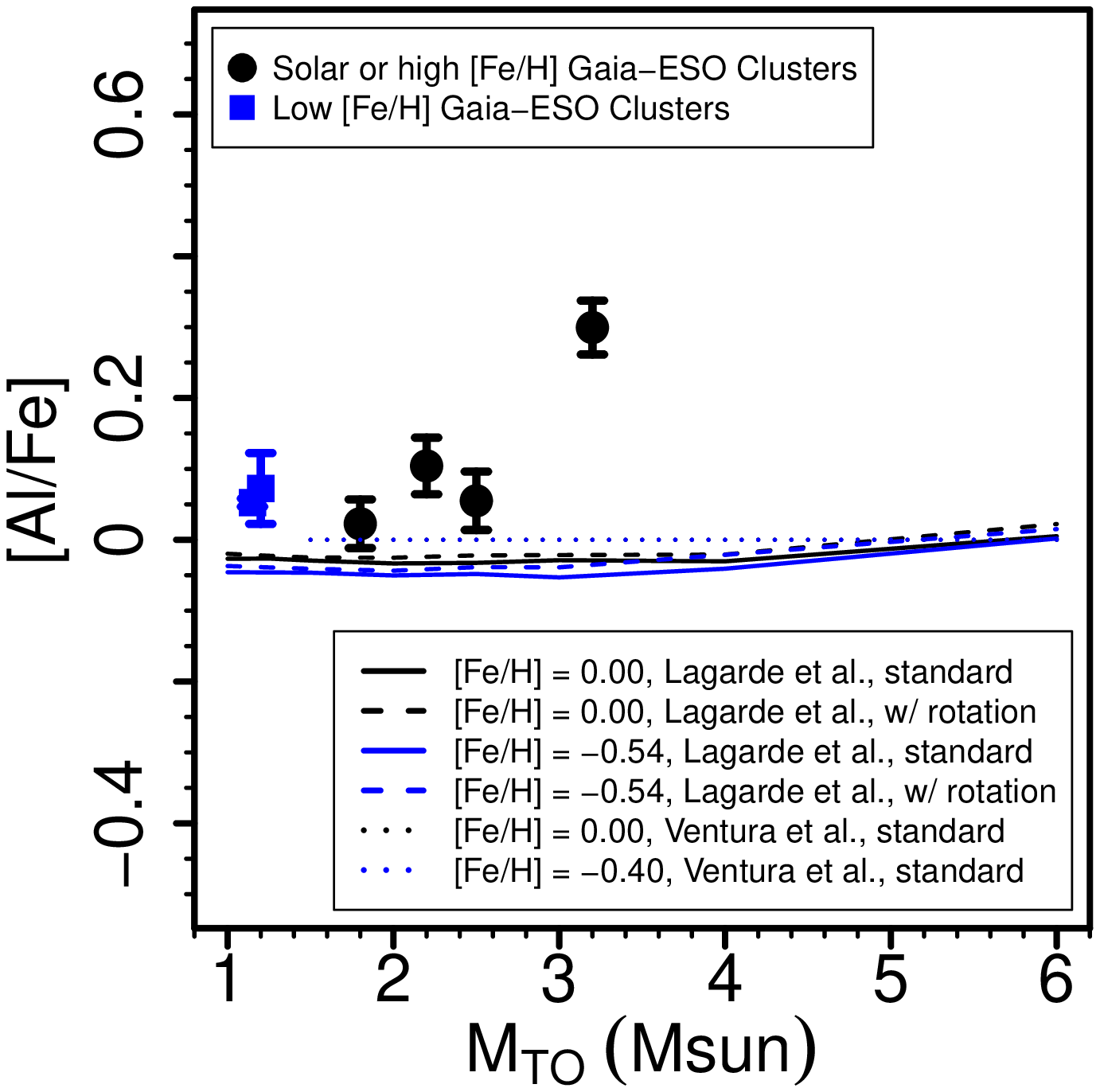}
 \caption{Mean cluster abundance, from giants only, after the selection of members and best-quality values. We estimate the uncertainty in the turn-off masses to be less than $\pm$ 0.1 $M_{\odot}$.}\label{fig:ocs}%
\end{figure*}

There is some discussion in the literature about overabundances of Na and Al in giants of open clusters and a possible connection with the first dredge-up \citep[see, e.g.,][and references therein]{2007AJ....134.1216J,2012MNRAS.422.1562S}. In this section, we revisit this issue using the Gaia-ESO sample of open cluster giants described above. The vast majority of the giants observed by Gaia-ESO are expected to be clump giants, and thus after completion of the first dredge-up.

We complemented the Gaia-ESO results with the Na abundance of clump giants in the Hyades open cluster determined by \citet{2012MNRAS.422.1562S}. The Na overabundance in the Hyades giants (age of $\sim$ 625 Myr and turn-off mass $\sim$ 2.58 $M_{\odot}$) was first found by \citet{1964ApJS....9...81H}. With well constrained atmospheric parameters (mostly independent of spectroscopy), and a critical selection of spectral lines, \citet{2012MNRAS.422.1562S} found [Na/Fe] = +0.30 in non-LTE \citep[also corrected using the grid of][]{2011A&A...528A.103L}. The adopted \emph{gf}\footnote{The product of the statistical weight \emph{g} of the lower energy level involved in the transition with the transition oscillator strength \emph{f} values of the Na lines} were the same as those used here, but accounting for differences in the solar reference abundances (but not in the stellar parameters scale), the Hyades have [Na/Fe] = +0.26 in the Gaia-ESO scale.

To compare with the observations, we use the evolutionary models computed by \citet{2012A&A...543A.108L} and \citet{2013MNRAS.431.3642V}. The models of \citet{2013MNRAS.431.3642V} include only convection as a mixing mechanism and were computed for two metallicities, solar and [Fe/H] = $-$0.40. The \citet{2012A&A...543A.108L} models were computed for solar metallicity and [Fe/H] = $-$0.54, and for the cases with and without rotation-induced mixing. The model with rotation also includes thermohaline mixing, but this process does not affect Na or Al. The initial rotation velocity of the modeled stars is 30\% of the critical velocity at the zero-age main sequence \citep[see][]{2014A&A...570C...2L}. If the initial rotation of the observed stars was different from that, the effect of rotation induced-mixing would also be different. Therefore, some scatter at a given mass can be expected, reflecting the scatter in the initial rotation of stars of the same mass. However, we do not have models computed with different initial rotation values and cannot judge the expected magnitude of such scatter. 

We do not renormalize the models to our adopted solar abundances. We consider both the observed and modeled [Na/Fe] and [Al/Fe] values to be relative values with respect to the abundances that the stars had during the main sequence. While this is strictly true for the models, as the stars had initially [Na/Fe] and [Al/Fe] = 0.0, for the observed giants the main-sequence Na abundances are unknown. For the effects of this discussion, we assume that the Sun is a good reference for the initial Na abundances of the stars, which by definition implies [Na/Fe] and [Al/Fe] = 0.0. An expanded discussion of dwarfs and giants abundances in a few open clusters will be possible with new Gaia-ESO observations, and will be the subject of a future paper.

\subsection{Model comparison with the Gaia-ESO sodium abundances}

As can be seen in the left panel of Fig.\ \ref{fig:ocs}, according to the stellar evolution models stars less massive than $\sim$ 1.5--2.0 $M_{\odot}$ do not change their Na surface abundance after the first dredge-up. For more massive stars, a change in the surface Na abundance is expected. In the \citet{2012A&A...543A.108L} models, the higher the stellar mass, the stronger the overabundance (for both models, with and without rotation). In the \citet{2013MNRAS.431.3642V} models, instead, the expected Na enhancement is constant above $\sim$ 3.0 $M_{\odot}$.

Regarding the observations, our results indicate that even when non-LTE effects are taken into account, some giants in open clusters still display Na overabundances with respect to the Sun. The observations also suggest an increase in Na enhancement as a function of stellar mass. However, with only one cluster beyond $\sim$ 3.0 $M_{\odot}$, we cannot state whether the Na overabundance continues to increase or reaches a plateau.

Because of the zero point uncertainty in [Na/Fe], we cannot exclude a small Na overabundance below $\sim$ 2 $M_{\odot}$. One cluster in particular, NGC 2243 with turn-off mass $\sim$ 1.2 $M_{\odot}$, displays a mild overabundance, [Na/Fe] = +0.10 $\pm$ 0.07 (average and standard deviation), although still in marginal agreement with the models within the errors. This is the most metal-poor cluster in our sample, [Fe/H] = $-$0.44. Nevertheless, the low-metallicity models for this mass range behave the same as the solar metallicity models, i.e., no Na overabundance is expected. 

For stars above $\sim$ 2 $M_{\odot}$ we can draw stronger conclusions. The Na overabundances are real, are not erased when non-LTE is taken into account, and they seem to increase with increasing stellar mass. The zero point uncertainty does not change this conclusion. As the Sun is a reference for all [Na/Fe] abundances, the values could move up or down in Fig.\ \ref{fig:ocs}, but the overabundances would not disappear completely and the trend with stellar mass would remain. 

In addition, the Na overabundances would likely remain in an analysis using more realistic three-dimensional (3D) model atmospheres. \citet{2007A&A...469..687C} and \citet{2013A&A...559A.102D} have compared Na abundances of giants derived using 1D and 3D model atmospheres, for a few representative cases. For the Na lines 6154 and 6160 \AA, with excitation potential $\sim$ 2 eV, the corrections are small ($\leq$ $\pm$0.05 dex) and could be positive (i.e., the 3D corrected Na abundances could be slightly larger than our values based on a 1D analysis).

Thus, we consider the trend in Fig.\ \ref{fig:ocs} real and a strong indication that the sodium overabundances in these stars are caused by internal evolutionary processes. In the future, new Gaia-ESO observations of giants in young clusters (age~$\sim$ 100 Myr; e.g., NGC 3532, NGC 6067, and NGC 6633) will help to further populate the high-mass end of Fig.\ \ref{fig:ocs}. This will help to expand the current discussion, and perhaps provide an opportunity to discriminate between models with and without rotation at the high-mass end. We note that, as reported in \citet{2015A&A...573A..55T}, the C and N abundances in clump giants of NGC 4815 and NGC 6705, and in both clump and evolved RGB stars in Trumpler 20 seem to agree better with models without rotation, although the models with rotation cannot be excluded because of their large error bars.

\subsection{Model comparison with the Gaia-ESO aluminium abundances}

The right panel of Fig.\ \ref{fig:ocs} suggests that below 3 $M_{\odot}$ the LTE abundance of Al in giants is constant around [Al/Fe] $\sim$ +0.06. Taking an average non-LTE correction into account on the order of $-$0.05 dex (Section \ref{sec:NLTE}), we find that the stars below 3 $M_{\odot}$ are consistent with [Al/Fe] = 0.00, i.e., no change in the surface abundance of Al after the first dredge-up. Thus, the observations agree well with the predictions of stellar evolution models. The small scatter in the observed abundances is consistent with the uncertainties. Even though there is some uncertainty in the zero point of our [Al/Fe] values (a maximum change of 0.07 dex), the lack of trend with stellar mass is a good indicator that there is no stellar evolutionary effect in the Al abundances. 

The only cluster above 3 $M_{\odot}$, NGC 6705, seems to have an enhanced Al abundance ([Al/Fe] = +0.30 dex in LTE), which would remain significant even after non-LTE corrections. However, we remark that stars in NGC 6705 seem to be $\alpha$-enhanced \citep[as discussed in][]{2014A&A...563A..44M,2015A&A...580A..85M}. While Al is not an $\alpha$-element, it does seem to behave as one, at least for metallicities between solar and [Fe/H] $\sim$ $-$1.0 (see Fig.\ \ref{fig:abunfeh}). We cannot discard the possibility that the $\alpha$-enhancement of NGC 6705 is accompanied by a similar Al enhancement. In fact, we note that the disk field stars analyzed by \citet{2014A&A...562A..71B} that have [Fe/H] $>$ 0.00 and [Mg/Fe] $>$ +0.1 are also enhanced in Al. Thus, the Al overabundance in NGC 6705 seems to be related to the environment where the cluster was formed. Indeed, \citet{2015A&A...580A..85M} made the hypothesis that NGC 6705 was enriched by a type II supernova in the mass range 15-18 $M_{\odot}$. The measurement of Al abundances in dwarfs of this cluster would help to clarify the situation, but in our sample Al abundances are only available for giants. New Gaia-ESO observations of giants in young clusters will also be useful in this context.

\subsection{Na enhancement: Literature results}\label{sec:lit}

\subsubsection{Open clusters}

We now check whether literature Na abundances support our conclusions above. For this, we take advantage of the compilation of Na abundances by \citet{2015MNRAS.446.3556M}. These authors conducted a homogenization of literature Na abundances in open cluster stars, changing the solar reference abundances and applying the non-LTE corrections of \citet{2011A&A...528A.103L}.

We extracted the Na abundances obtained only from the analysis of giants from their Table 2. This included a total of eleven open clusters, but we further excluded NGC 6791. For this cluster, the compilation listed the Na abundances from \citet{2012ApJ...756L..40G}. These authors claimed to observe a Na-O anticorrelation similar to the anticorrelation common in globular clusters. We do not include these results to avoid introducing a different physical effect in the discussion. We also remark that the Na-O anticorrelation in NGC 6791 was not confirmed by both \citet{2014ApJ...796...68B} and \citet{2015ApJ...798L..41C}, and that \citet{2015ApJ...799..202B} did not find any spread of oxygen abundances in turn-off stars of the cluster.

\begin{figure}
\centering
\includegraphics[height = 8cm]{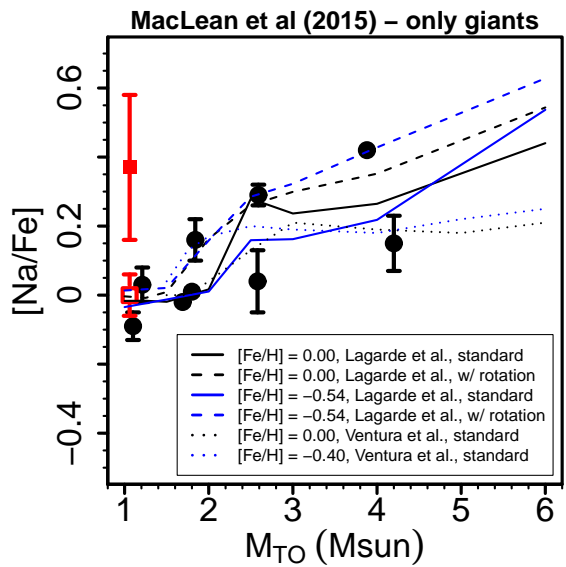}
\caption{Mean [Na/Fe], in non-LTE, only of giants for clusters in the compilation of {\citet{2015MNRAS.446.3556M}.} The red solid square is the original [Na/Fe] value of Collinder 261 in that compilation, while the red open square is our revised value as discussed in the text. }\label{fig:maclean}%
\end{figure}

Figure \ref{fig:maclean} shows the [Na/Fe] ratios extracted from \citet{2015MNRAS.446.3556M} as a function of the turn-off mass of the clusters. Ages and turn-off masses for the ten clusters (i.e., Berkeley 39, Collinder 261, Hyades, IC 4651, M 67, NGC 3114, NGC 6134, NGC 6475, NGC 7789, and Trumpler 20) were taken from a variety of references. These include some values of turn-off masses that we adopted in this work (e.g., for the Hyades and Trumpler 20) and those quoted in the original sources of the abundances \citep{2000A&A...360..499T,2005A&A...431..933T,2003AJ....126.2372F,2005A&A...441..131C,2009ApJ...701..837S,2009A&A...504..845V,2010MNRAS.407.1866M,2011MNRAS.413.2199M,2012A&A...548A.122B,2013A&A...554A...2S,2014A&A...562A..39C}. Apart from one cluster, Collinder 261 at 1.06 $M_{\odot}$ and [Na/Fe] = +0.37, Fig.\  \ref{fig:maclean} shows a trend of [Na/Fe] with stellar mass similar to that seen in our own sample. The scatter seems to be larger, but we note that this is a compilation of literature results. In addition, we do not know for certain the evolutionary status of all these giants. Some might be before the end of the first dredge-up. Thus we refrain from over interpreting the scatter and postpone a more detailed discussion for when a larger sample of homogeneous Gaia-ESO results become available.

The abundances of Collinder 261 are originally from \citet{2003AJ....126.2372F}, [Fe/H] = $-$0.22 and [Na/Fe] = +0.48 $\pm$ 0.22 (in LTE); and \citet{2005A&A...441..131C}, [Fe/H] = $-$0.03 and [Na/Fe] = +0.33 $\pm$ 0.06 \citep[in non-LTE, with corrections from][]{1999A&A...350..955G}. Correcting these values for the \citet{2015MNRAS.446.3556M} solar scale, we obtain [Na/Fe] = +0.55 and [Na/Fe] = +0.26, respectively. \citeauthor{2005A&A...441..131C} discussed the difference between the two results listing as possible reasons, for example, the higher spectral resolution of their own data and their more robust determination of microturbulence. We thus prefer to adopt the \citeauthor{2005A&A...441..131C} analysis as the reference Na abundance for Collinder 261.

The Na non-LTE correction of \citet{1999A&A...350..955G} for a star of $T_{\rm eff}$ = 4000 K and $\log~g$ = 1.5 dex is on the order of +0.20 dex\footnote{Online table available at Vizier: \url{http://vizier.cfa.harvard.edu/viz-bin/VizieR-3?-source=J/A\%2bA/350/955/abundcor}}. As discussed before, the average correction from the \citet{2011A&A...528A.103L} grid, however, is on the order of $-$0.10 dex. Taking this difference into account, the non-LTE Na abundance for Collinder 261 would instead be [Na/Fe] $\sim$ 0.00.

With this revised value, the Na abundance in Collinder 261 turns out to be in excellent agreement with the model expectations for its turn-off mass. These literature results support the trend of increasing Na overabundance with increasing stellar mass and, moreover, also suggest that there is no significant Na overabundance for low-mass stars below 2.0 $M_{\odot}$. This supports the idea that evolutionary mixing processes are the origin of the observed Na enhancements in giants with M $\gtrsim$ 2.0 $M_{\odot}$.

\begin{figure}
\centering
\includegraphics[height = 8cm]{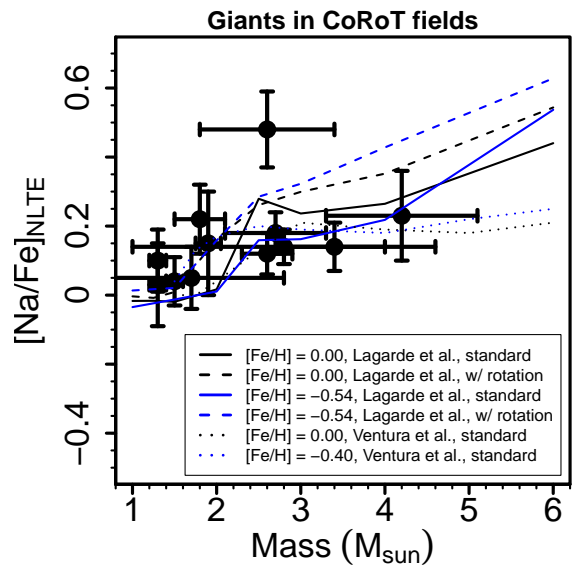}
\caption{Sodium abundances as a function of stellar mass for a sample of literature giants with seismic data.}\label{fig:corot}%
\end{figure}

\subsubsection{Giants with seismic masses}

It is not straightforward to look for the [Na/Fe] vs.\ stellar mass trend in field giants because accurate masses and evolutionary stages are notoriously difficult to determine for field stars. This has started to change with the advent of asteroseismic space missions, such as CoRoT \citep[Convection, Rotation, and planetary Transits,]{2006cosp...36.3749B} and Kepler \citep{2010Sci...327..977B}. By studying the oscillation properties of giants it has become possible to estimate their masses, among other quantities \citep[see e.g.,][]{2008ApJ...674L..53S,2010A&A...509A..77K,2010A&A...517A..22M}.

Taking advantage of this, we extracted a sample of 16 giants with LTE Na abundances from \citet{2014A&A...564A.119M}, which are determined using a set of atmospheric parameters including $\log~g$ based on the ionization equilibrium. Seismic masses for the same giants were adopted from \citet{2015A&A...580A.141L}. The giants have metallicities between [Fe/H] = $-$0.35 and +0.13. Twelve of these giants have been observed by CoRoT, while the other four are bright well-studied giants with asteroseismic data available from elsewhere and used as reference stars by \citet{2014A&A...564A.119M}. We computed non-LTE Na abundance corrections again using the grids computed by \citet{2011A&A...528A.103L}.

The [Na/Fe] ratios of these giants are shown as a function of stellar mass in Fig. \ref{fig:corot}. This data set corroborates our previous conclusions. A similar trend between [Na/Fe] and stellar mass is seen here. Stars more massive than $\sim$ 2.0-2.5 $M_{\odot}$ have on average higher Na enhancement than stars less massive than that. The one outlier with higher [Na/Fe] than expected by the models is \object{HD 50890}. This star, however, has unusually large line broadening \citep{2014A&A...564A.119M} and, according to the more detailed analysis of \citet{2012A&A...538A..73B} using the same CoRoT data, could have a mass of up to 5 $M_{\odot}$. The higher mass would improve the agreement with the models.
\begin{figure*}
\centering
\includegraphics[height = 7cm]{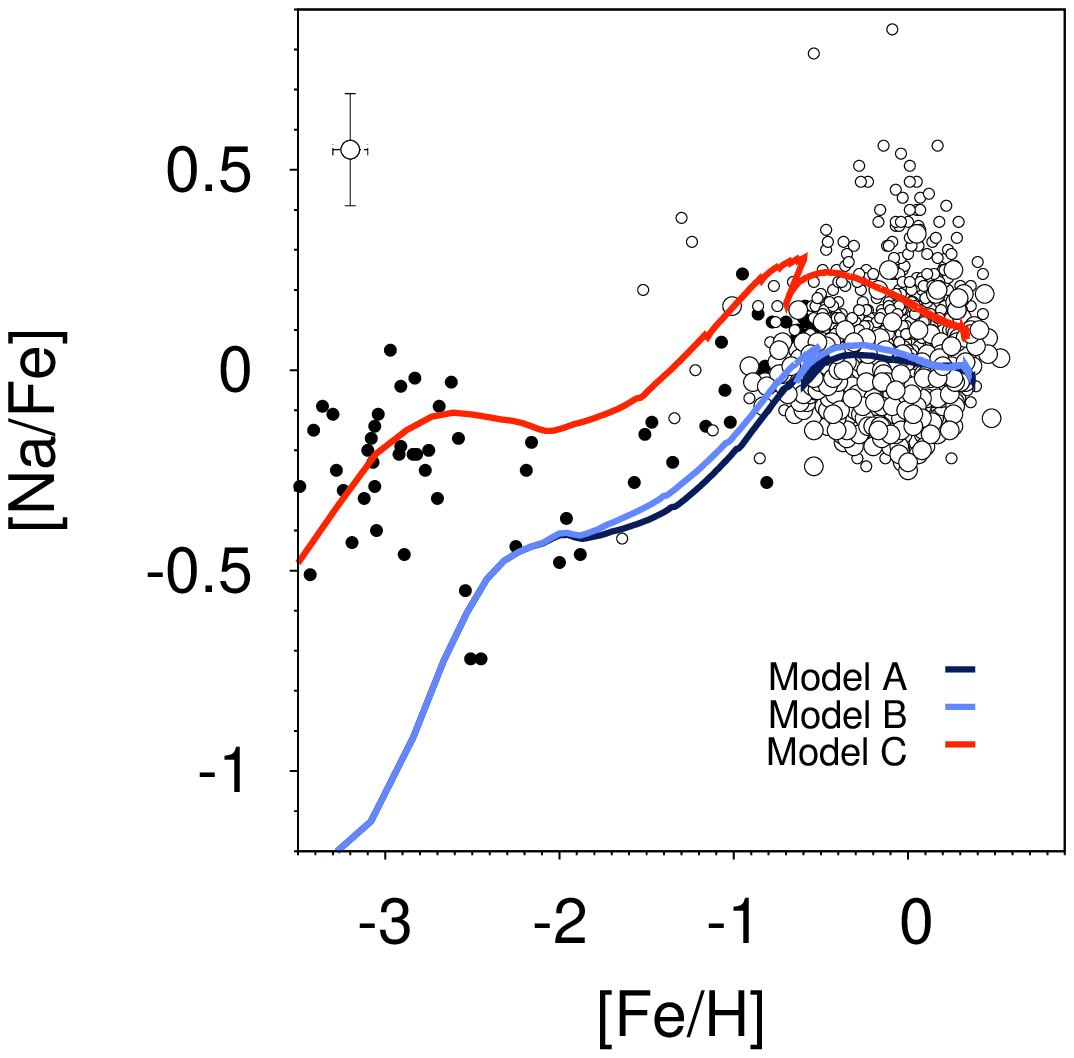}
\includegraphics[height = 7cm]{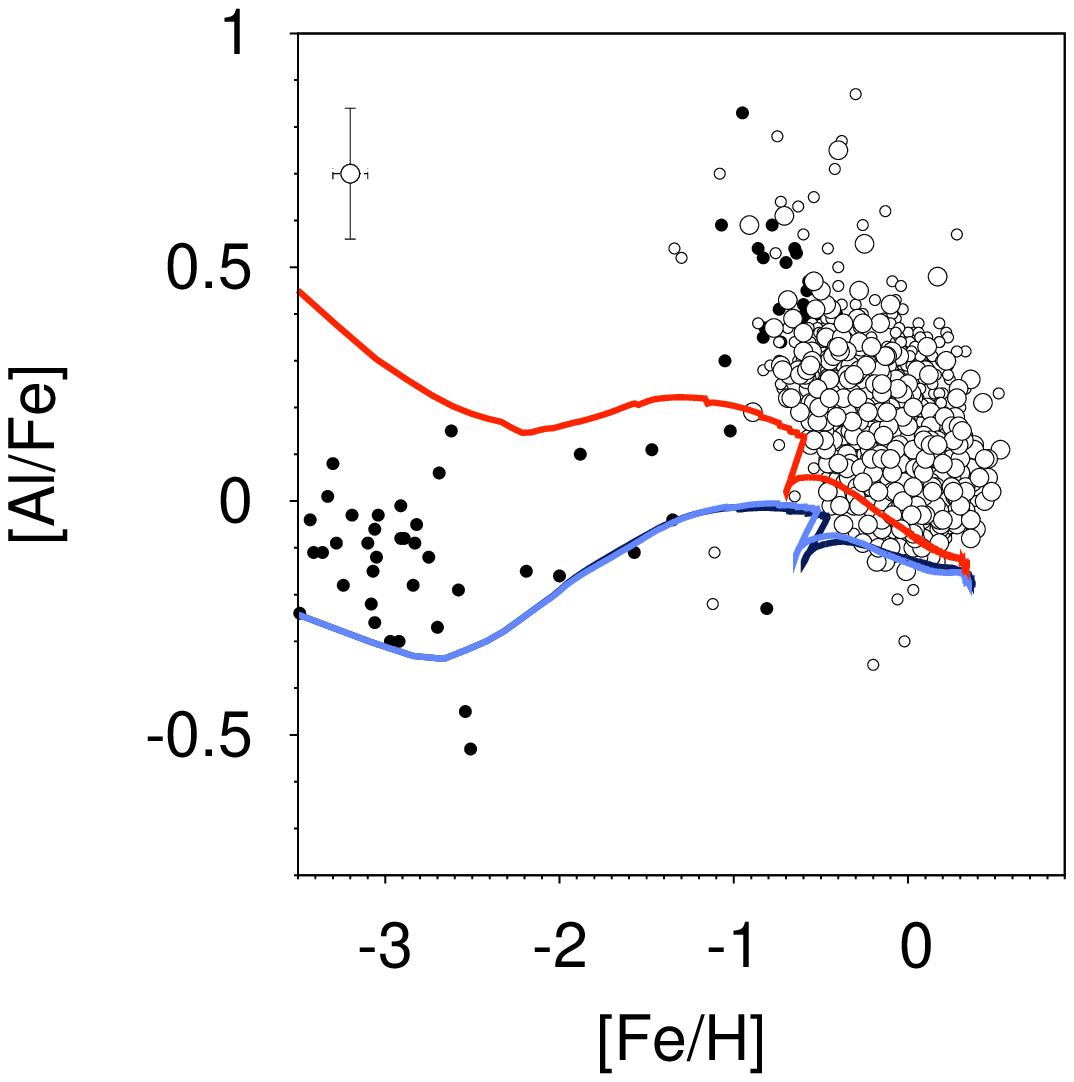}
 \caption{Runs of [Na/Fe] \emph{(left panel)} and [Al/Fe] \emph{(right panel)} with [Fe/H] predicted by chemical evolution models for the solar neighborhood adopting different stellar yields (see text; legend on the lower right corner of the left panel). Data for low-metallicity stars (filled circles) are from \citet{2006A&A...451.1065G}, \citet[][for Na only]{2007A&A...464.1081A}, and \citet[][for Al only]{2008A&A...481..481A}. High-quality data for our sample dwarfs (selected as in Sect.\ \ref{sec:best}) are shown as large empty circles, while the small empty circles refer to the full sample, including lower quality data and giants. All measured Na abundances were corrected for non-LTE effects. Typical error bars are $\sim$0.15 dex for [El/Fe] and $\sim$0.10 dex for [Fe/H] in this and all following plots. A typical error bar ($\pm$ 0.10 dex for [Elem./H] and $\pm$ 0.14 dex for [Elem./Fe] is shown in the upper left of each panel.)}\label{fig:evofeh}%
\end{figure*}
%

\section{Galactic chemical evolution of Na and Al}\label{sec:chemical}

The history of Na and Al enrichment on a Galactic scale is not well understood yet. Chemical evolution models adopting different stellar yields can reproduce satisfactorily well the average trend of either [Na/Fe] vs. [Fe/H] or [Al/Fe] vs. [Fe/H] in the solar vicinity, but can never reproduce both simultaneously \citep[see e.g.][their figures~22 and 10, respectively]{2010A&A...522A..32R,2013ARA&A..51..457N}. Furthermore, the increase of [Na/Fe] with metallicity observed for [Fe/H]~$>$ 0.00 is not explained by the models. 

\subsection{Trends with metallicity}

In Fig.~\ref{fig:evofeh}, we show the predictions of model~15 of \citet[][]{2010A&A...522A..32R}; labeled Model~A here, compared to those of other two models, obtained by assuming up-to-date prescriptions about stellar nucleosynthesis:

\begin{itemize}
\item Model~A adopts the yields by \citet{2010MNRAS.403.1413K} for low- and intermediate-mass stars and the yields by \citet{2006ApJ...653.1145K} for massive stars; in particular, it assumes that all stars above 20~M$_\odot$ explode as hypernovae, with energies much larger than normal supernovae;
\item Model~B is the same as Model~A, but the yields for low- and intermediate-mass stars are from recent work published in \citet{2013MNRAS.431.3642V,2014MNRAS.437.3274V,2014MNRAS.439..977V} and extend to super-solar metallicities;
\item Model~C is the same as Model~B, but all massive stars explode as core-collapse supernovae with energies on the order of $~$10$^{51}$~ergs.
\end{itemize}

The model predictions are compared to measurements of Na and Al for dwarf stars in our sample (to avoid mixing effects on abundances) with high-quality data (dispersion below 0.15 dex and results based on four or more pipelines; see Section~\ref{sec:best}; large empty circles at [Fe/H]~$\ge -$1.0 dex); adding giants and lower quality data (small empty circles at [Fe/H]~$\ge -$1.5 dex) increases the dispersion, as expected. The data for the halo (turnoff and giant stars; small filled circles) are from \citet{2006A&A...451.1065G}, \citet[][for Na only]{2007A&A...464.1081A} and \citet[][for Al only]{2008A&A...481..481A}; to minimize spurious effects due to mass transfer from companions in binary systems or stellar evolution, we do not show the abundances of either known carbon-rich stars or mixed giants \citep[i.e., stars located after the RGB bump; see][and references therein]{2007A&A...464.1081A}. All the ratios are normalized to the reference solar abundances adopted in this work (see Section~\ref{sec:sun}). Furthermore, all Na abundances are corrected for non-LTE effects (see Section~\ref{sec:NLTE}).

Notwithstanding the use of updated stellar yields, the detailed runs of [Na/Fe] and [Al/Fe] with [Fe/H] in the solar vicinity remain largely unexplained. As expected (see Introduction), the contribution to Na and Al production from low- and intermediate-mass stars is negligible on a Galactic scale. The assumption that all stars above 20~M$_\odot$ explode as hypernovae (Models~A and B) results in the lowest theoretical [Al/Fe] ratios at the lowest metallicities, which is in reasonably good agreement with the observations, but also leads to extremely low [Na/Fe] ratios that do not match Na observations for [Fe/H]~$< -$2.5. The model adopting the yields by \citet{2006ApJ...653.1145K} for normal core-collapse supernovae (Model~C) does a better job for Na for [Fe/H]~$< -$2.5, but severely overestimates the [Al/Fe] ratios in the halo. Moreover, while it explains qualitatively the decreasing trend of [Al/Fe] with metallicity for [Fe/H]~$>-$1.0, it underproduces Al in the disk overall.

\begin{figure*}
\centering
\includegraphics[height = 7cm]{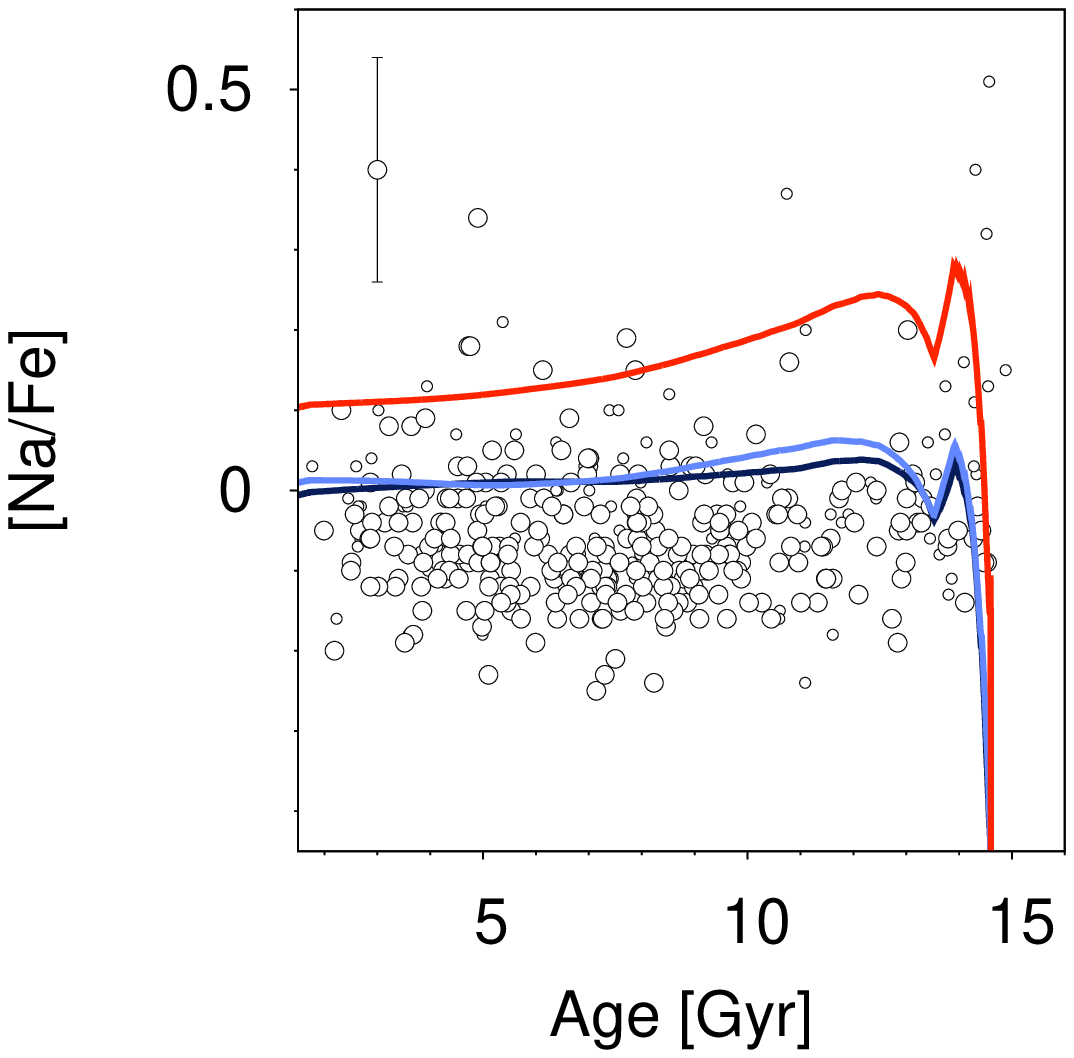}
\includegraphics[height = 7cm]{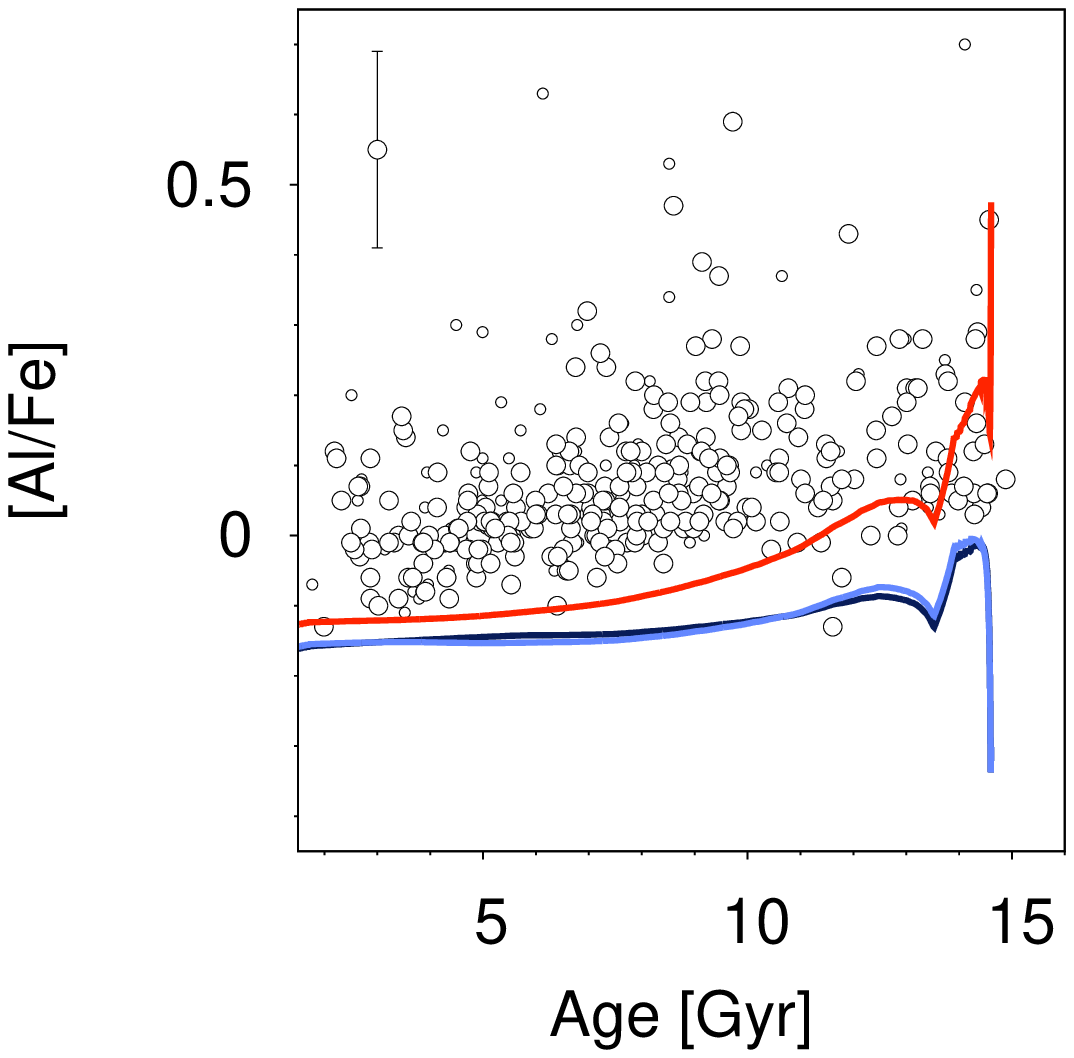}
 \caption{Runs of [Na/Fe] \emph{(left panel)} and [Al/Fe] \emph{(right panel)} against age. Ages are available only for a subsample of Gaia-ESO solar neighborhood dwarfs with [Fe/H] $> -$1.0. Models and error bars are the same as in Fig.\ \ref{fig:evofeh}.}\label{fig:evoage}%
\end{figure*}
\begin{figure*}
\centering
\includegraphics[height = 5.4cm]{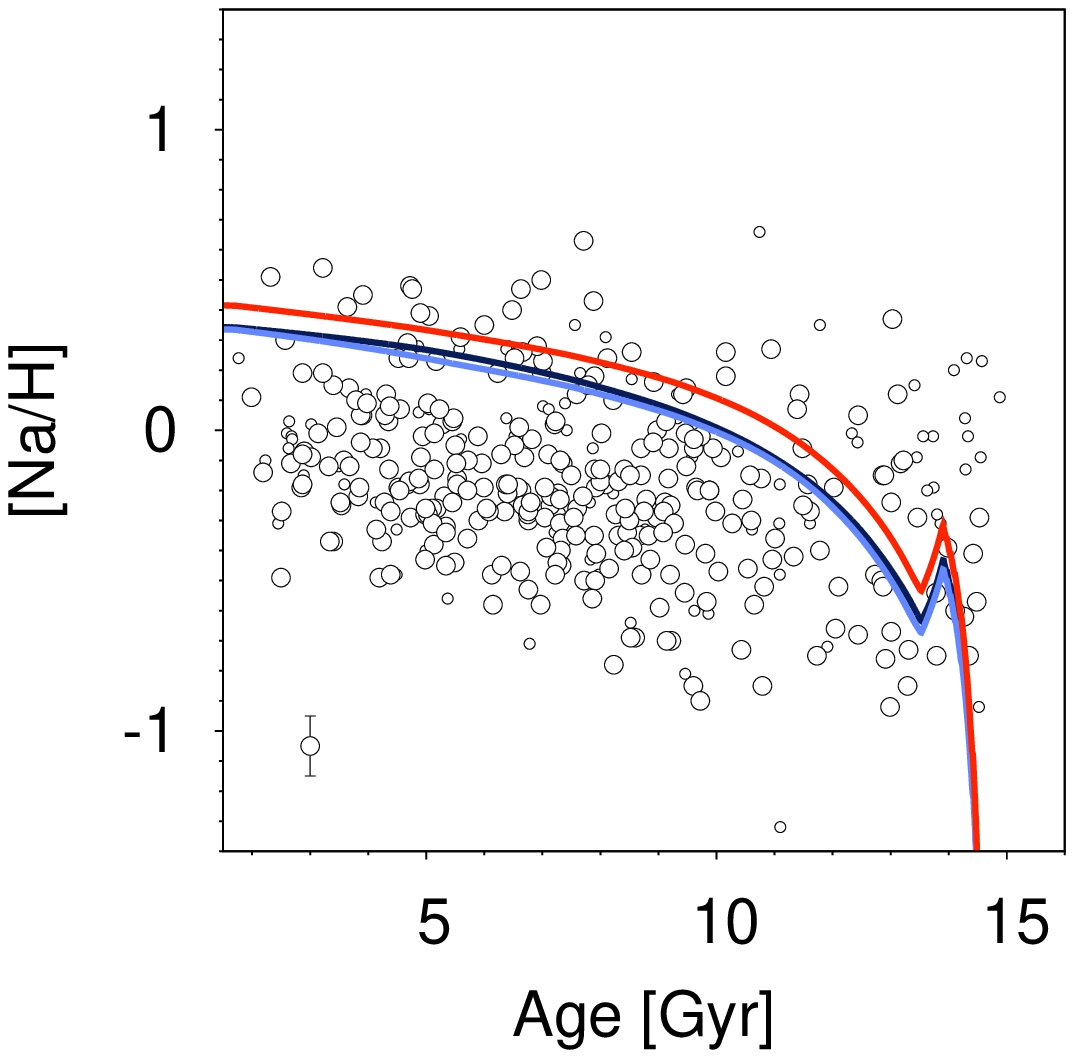}
\includegraphics[height = 5.4cm]{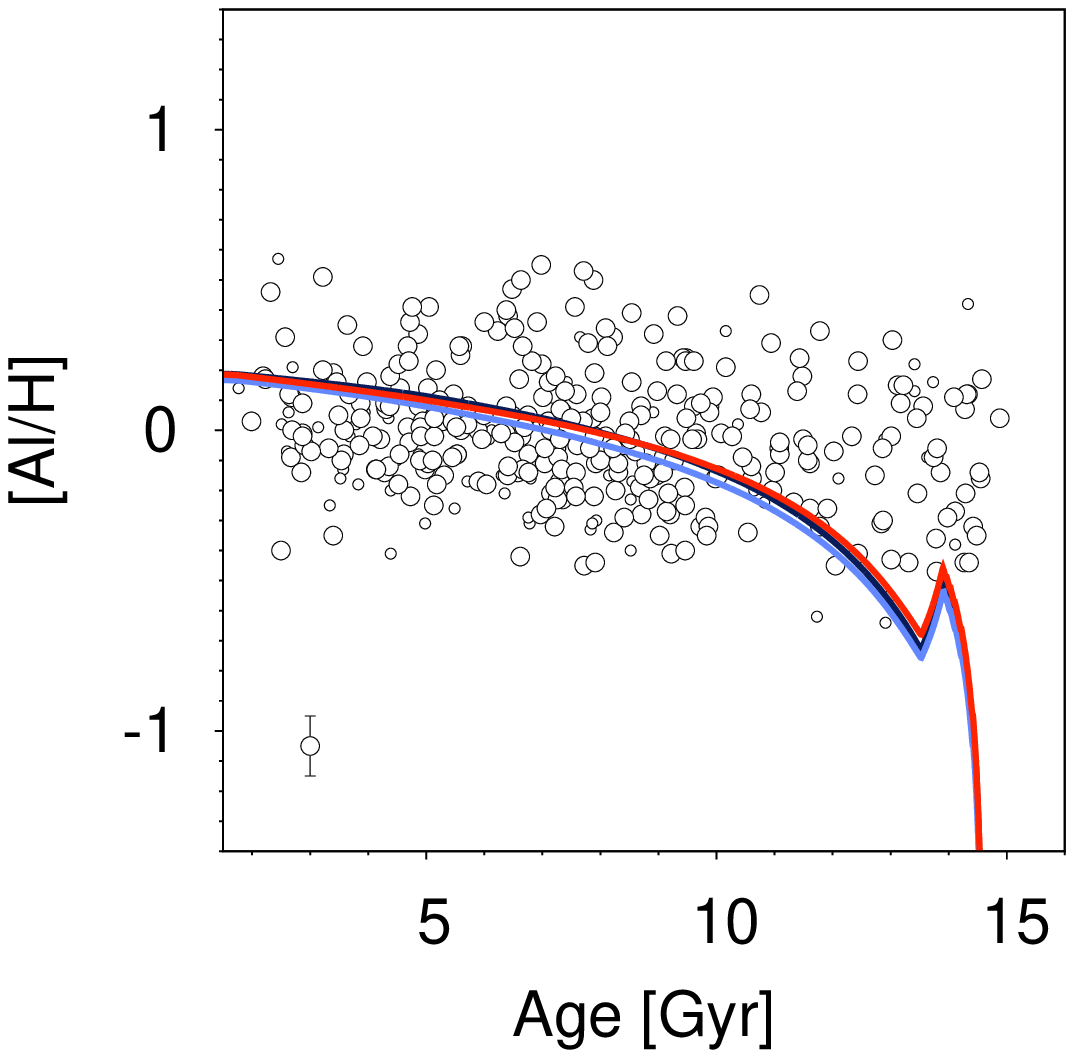}
\includegraphics[height = 5.4cm]{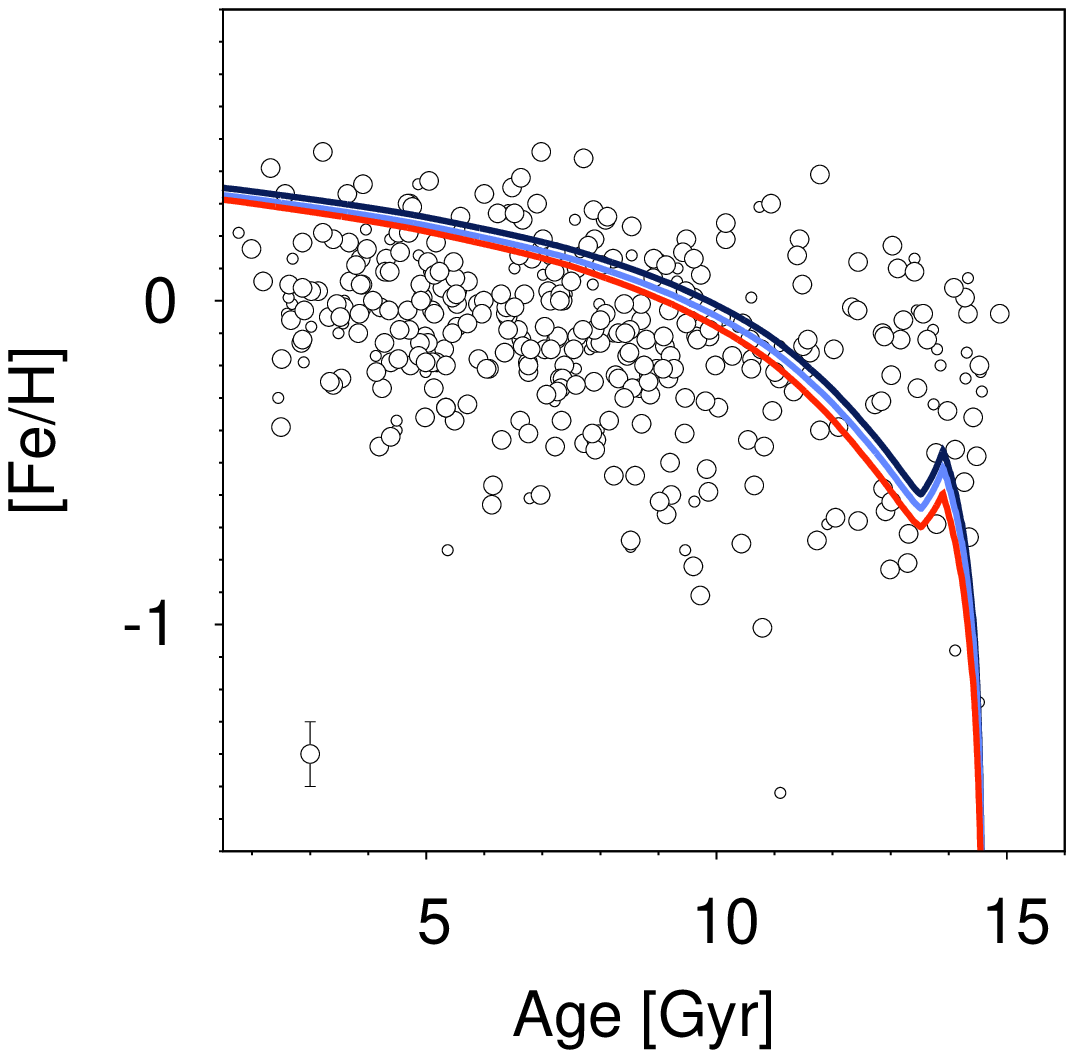}
 \caption{Comparison of the observed ratios [Na/H] (\emph{left panel}), [Al/H] (\emph{central panel}) and [Fe/H] (\emph{right panel}) and the model predictions as a function of age. Models and error bars are the same as in Fig.\ \ref{fig:evofeh}.}\label{fig:agenahfeh}%
\end{figure*}
\begin{figure*}
\centering
\includegraphics[height = 7cm]{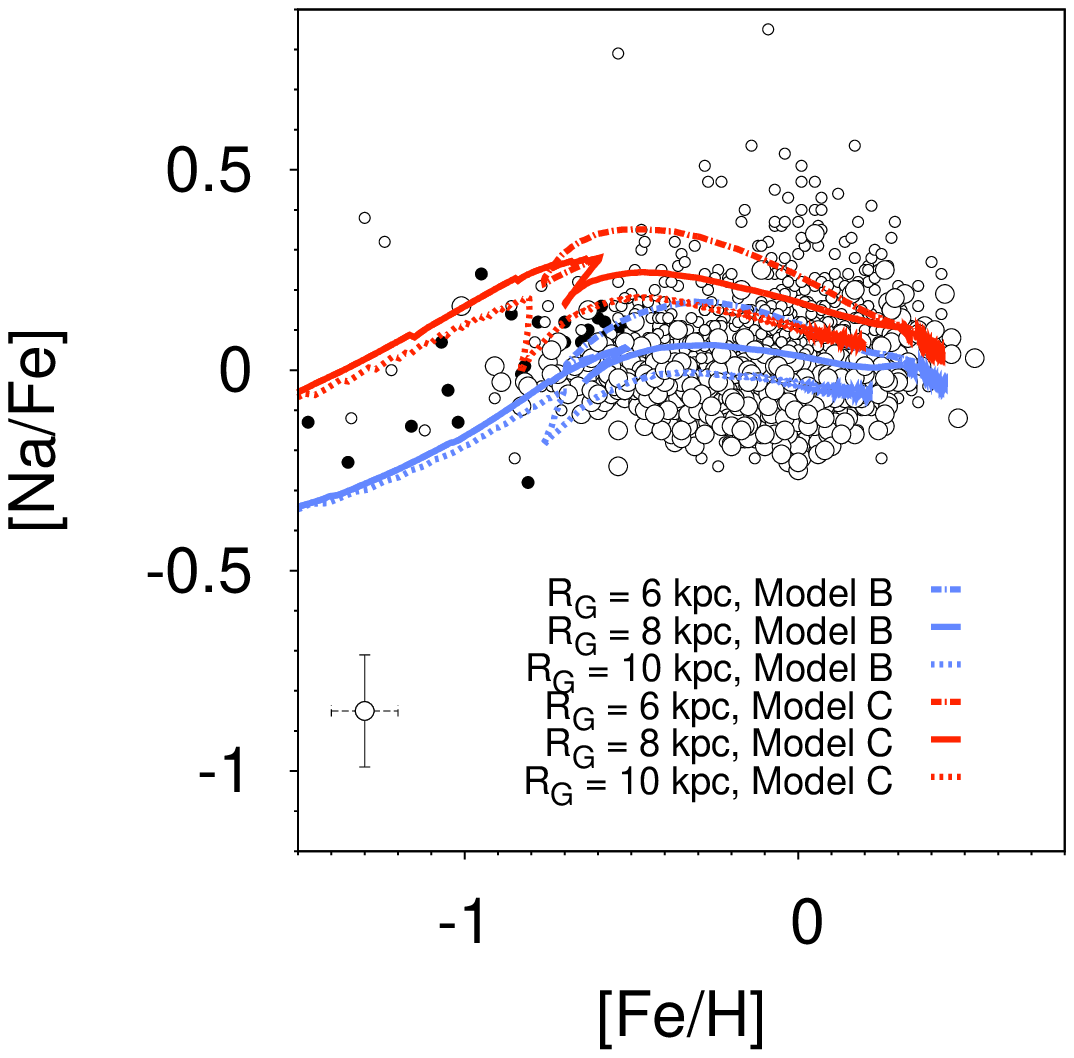}
\includegraphics[height = 7cm]{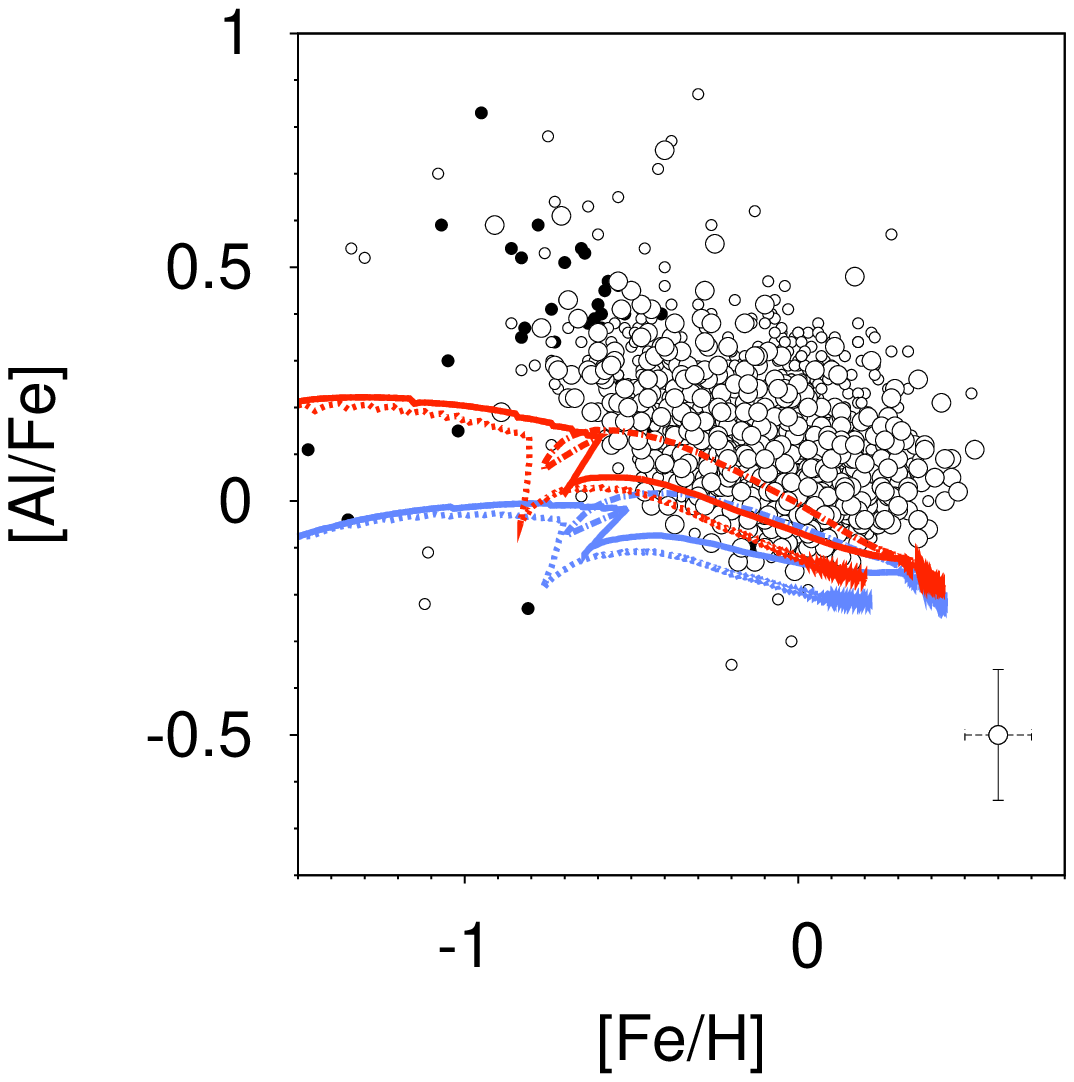}
 \caption{Data for [Na/Fe] vs [Fe/H] and [Al/Fe] vs [Fe/H] (\emph{left} and \emph{right panels,} respectively) compared to the predictions of Models~B and C run at R$_{\rm G}$ = 6, 8, and 10 kpc. Error bars are the same as in Fig.\ \ref{fig:evofeh}}\label{fig:evofehR}%
\end{figure*}

None of the models can explain the observed increase of [Na/Fe] for [Fe/H]~$>$ 0, which could suggest that the models lack some site of Na production at later stages. The missing source should be sufficiently strong to reverse the decreasing trend of [Na/Fe] versus [Fe/H] due to the delayed Fe production from supernovae (SNe) Ia. For this late Na production, we can think of two possible sites, i.e., SNe Ia and novae. The models shown here already account for some Na production from SNe Ia with yields from \citet{1999ApJS..125..439I}, but the contribution is negligible. Novae might also act in the right direction, since they restore the products of explosive H burning on relatively long timescales \citep[see, e.g.,][and references therein]{1999A&A...352..117R,2003MNRAS.342..185R}. Indeed, novae have been shown to be able to contribute important amounts of $^{7}$Li, $^{13}$C, $^{15}$N, $^{17}$O, $^{22}$Na, and $^{26}$Al \citep{1998ApJ...494..680J,2007JPhG...34..431J}, although the actual yields remain highly uncertain. It is worth remarking that \citet{2015ApJ...808L..14I} detected Li expelled by a nova system \citep[see also][]{2015Natur.518..381T} and found a total amount of Li ejected in a single nova outburst that is significantly larger than expected from hydrodynamic nova models by \citet{1998ApJ...494..680J}. In this context, it would be interesting to investigate whether some extra $^{23}$Na can be produced in these same events in amounts that would reverse the decreasing trend of [Na/Fe] with time predicted by current chemical evolution models (see next section). On the other hand, the possibility exists that the explanation of the increasing trend of [Na/Fe] with time lies elsewhere. Indeed, [Ni/Fe] also shows a clear upturn at [Fe/H] $>$ 0 \citep{2014A&A...562A..71B} which is hardly attributable to nova nucleosynthesis.

\subsection{Trends with age}

In Fig.\ \ref{fig:evoage}, we show the plots of [Na/Fe] (left panel) and [Al/Fe] vs. age (right panel). Data are only available for a subset of the solar neighborhood Gaia-ESO dwarf sample with [Fe/H] $\geq -$1.0. The model predictions refer to the evolution of a thin-disk component (ages younger than 13.5 Gyr in Fig.\ \ref{fig:evoage}), plus a thick-disk and halo component (ages older than 13.5 Gyr). Since the chemical evolution model assumes an age of the Universe of 13.7 Gyr and the procedure outlined in Sect.~\ref{sec:ages} leads to age estimates as old as $\sim$15 Gyr, we scaled the chemical evolution model results to an age of the Universe of 15 Gyr. In principle, comparing the observed and theoretical trends with age is quite instructive and might help to indicate why the models fail to reproduce the full behavior of the abundances. 

Figure \ref{fig:evoage} shows that the observed trend of the [Na/Fe] ratio in thin-disk stars is basically flat with respect to age. On the other hand, the models seem to predict (slightly) higher [Na/Fe] values for older stars. It is worth stressing at this point that reliable stellar ages could be derived only for half of the original Gaia-ESO sample (see Section~\ref{sec:ages}); therefore, our view of Na evolution could be biased in the abundance-versus-age diagram. Indeed, we notice that only a minority of the stars that have [Na/Fe]$>$0.00 in the [Na/Fe] vs. [Fe/H] plot (Fig.\ \ref{fig:evofeh}, left panel) is found in the [Na/Fe] vs. age plot (Fig.\ \ref{fig:evoage}, left panel). If most of the stars with [Na/Fe]$>$0.00 were young in age, the disagreement between the predicted and observed trends would worsen and this would strongly point to a missing late Na source. For Al, in the solar neighborhood thin-disk stars there seems to be a weak trend of decreasing [Al/Fe] for younger stars. As also seen in the trends with [Fe/H], the models seem to predict consistently lower [Al/Fe] than what is observed.

To separate the effects of the Na and Al evolution from that of Fe, we plot in Fig. \ref{fig:agenahfeh} the trends of [Na/H], [Al/H], and [Fe/H] as a function of age (left, middle, and right panels, respectively). The models seem to predict an increasing trend for [Na/H] that agrees with the observations but, again, we note that some Na-rich stars could be missing from the plot. Also, the increase in [Fe/H] predicted by the model is a bit steeper than the increase suggested by the observations. As for aluminium, there is a tendency of models to underproduce this element during the early disk evolution, while the predictions agree with the average observed trend of increasing [Al/H] for ages younger than 7 Gyr (Fig.\ \ref{fig:agenahfeh}, central panel). The stellar yields for this element clearly need to be revised as well.

\subsection{Trends with galactocentric distance}

It is well known that stars in a galactic disk can undergo important radial displacements and that these radial motions influence the chemical properties of the stellar populations \citep[see][among others]{2002MNRAS.336..785S,2008MNRAS.388.1175H,2008ApJ...684L..79R,2009MNRAS.396..203S,2013A&A...558A...9M,2013MNRAS.436.1479K}. The migrating stars, in fact, coming from different Galactic regions, bear the imprints of different evolutionary rates. This has been put forward as a likely explanation for most of the observed spread in the age metallicity and [El/Fe] vs. [Fe/H] relations of solar neighborhood stars some 20 years ago \citep[][and references therein]{1993A&A...280..136F}.

Our subsample of Gaia-ESO solar neighborhood dwarfs with reasonable age estimates confirms previous findings, showing that the dispersion in [Fe/H] values increases with increasing age. The situation for the ratios involving Na and Al is less clear (cf. Figs.\ \ref{fig:evoage} and \ref{fig:agenahfeh}). A detailed discussion of the amount and significance of the spreads is beyond the aim of this paper. However, in Fig.\ \ref{fig:evofehR} we compare the [Na/Fe] and [Al/Fe] ratios of Gaia-ESO field dwarfs as functions of [Fe/H] to the trends predicted by Models~B and C (namely, with and without hypernovae, respectively) at different Galactic radii, $R_{\mathrm{G}} =$~6, 8, and 10~kpc. The models assume a star formation efficiency that varies with radius after \citet[][see their figure~1]{2015ApJ...802..129S}. The dispersion in the abundance ratios could be explained, at least partly, by the radial migration of stars that formed at different radii and ended up in the solar vicinity. However, since we are using a pure chemical evolution model that does not include a detailed treatment of the stellar motions, we cannot make any quantitative prediction about the fractions of stars that are expected to be born at different radii.

%

\section{Summary}\label{sec:end}

We used new Na and Al abundances determined within the Gaia-ESO Survey, to readdress the behavior of these elements in what concerns both stellar and Galactic chemical evolution. For the stellar evolution discussion, we used a sample of giants in six open clusters, ranging in age from 300 Myr to 4.5 Gyr. For the chemical evolution discussion, we used a sample of $\sim$ 600 solar neighborhood dwarfs, complemented by halo stars from the literature. The Na abundances were corrected for non-LTE effects, and no corrections were applied to the Al abundances.

The average non-LTE Na abundances of the cluster giants show a trend of increasing [Na/Fe] with increasing stellar mass, which is in agreement with expectations of stellar evolution models. Similar trends are seen in a selection of literature Na abundances of open cluster giants and in field giants with seismic masses derived thanks to CoRoT light curves. We consider the trend with mass as strong evidence of the stellar evolution origin of the surface Na enhancement seen in these giants. Nevertheless, for stars with mass below $\sim$ 2 $M_{\odot}$, we cannot exclude a small Na enhancement, in disagreement with model predictions because of remaining uncertainties and possible systematics in the abundances. For stars with mass above $\sim$ 2 $M_{\odot}$, we are not able to differentiate between models with and without rotation induced mixing.

Regarding Al, no convincing evidence for a trend of [Al/Fe] with stellar mass was found. Below $\sim$ 3 $M_{\odot}$, the giants in our sample show a constant Al abundance. The only cluster with enhanced Al abundance, NGC 6705 with turn-off mass above 3 $M_{\odot}$, has peculiar chemical composition. This suggests that its Al enhancement has origin in the environment where the cluster was formed \citep[see][]{2015A&A...580A..85M}. 

The disagreement between Galactic chemical evolution models and observations for Na and Al in the solar neighborhood remains, even with the use of up-to-date stellar yields. The explanation for this disagreement does not seem to lie in low- and intermediate-mass stars, as their contribution to the increase in Na and Al in the Galaxy seems to be negligible. The average trend of [Na/Fe] with [Fe/H] in solar neighborhood dwarfs can be reproduced apart from the increase of [Na/Fe] at higher metallicities. The observed and predicted trend of [Na/H] with stellar age is, in principle, instructive to indicate where the problems seem to be. We note, however, that most of the stars with [Na/Fe]$>$0.00 have no reliable age determinations. This makes them disappear in the abundance vs. age diagram, which strongly affects any conclusion we might draw from this plot. Based on the [Na/Fe] vs. [Fe/H] diagram, we speculate that it is likely that some significant site of late Na production is missing from the models.

The failure to reproduce the behavior of the Al abundances with metallicity and age is even more striking. For the solar neighborhood dwarfs with [Fe/H] $> -$ 1.0, Al is underproduced at all ages, but the youngest ones. For halo stars of lower metallicity, the models that nicely reproduce Na observations consistently overproduce Al. Clearly, a better understanding of the nucleosynthesis of Al is needed.

We will revisit the remaining open issues in our analysis when the Gaia-ESO survey is complete (it is currently in its fourth year of observations). On the stellar evolution side, further comparisons between low-mass dwarfs and giants in the same open clusters are needed to better understand possible differences in Na abundances. The increase in the sample of younger clusters will also facilitate a better discussion of Na and Al in giants with masses above 3 $M_{\odot}$. On the chemical evolution side, we expect abundances with smaller uncertainties for a larger sample of solar neighborhood dwarfs and a new set of yields for massive stars computed considering the effects of stellar rotation on nucleosynthesis, from the pre-main sequence up to the explosive stages \citep[M. Limongi, private communication, and][]{2015IAUS..307....1C}. This will provide tighter constraints on the evolution of Na and Al in the Galactic disk.  

\begin{acknowledgements}
R.S. acknowledges support by the National Science Center of Poland through grant 2012/07/B/ST9/04428. T.M. acknowledges financial support from Belspo for contract PRODEX GAIA-DPAC. V.A. acknowledges the support from the Funda\c{c}\~ao para a Ci\^encia e a Tecnologia (FCT) in the form of the grants SFRH/BPD/70574/2010 and PTDC/FIS-AST/1526/2014. G.T. acknowledges support by the grant from the Research Council of Lithuania (MIP-082/2015). P.F. acknowledges support from the CNES. U.H. acknowledges support from the Swedish National Space Board (SNSB). Based on data products from observations made with ESO Telescopes at the La Silla Paranal Observatory under programme ID 188.B-3002. These data products have been processed by the Cambridge Astronomy Survey Unit (CASU) at the Institute of Astronomy, University of Cambridge, and by the FLAMES/UVES reduction team at INAF/Osservatorio Astrofisico di Arcetri. These data have been obtained from the Gaia-ESO Survey Data Archive, prepared and hosted by the Wide Field Astronomy Unit, Institute for Astronomy, University of Edinburgh, which is funded by the UK Science and Technology Facilities Council. This work was partly supported by the European Union FP7 programme through ERC grant number 320360 and by the Leverhulme Trust through grant RPG-2012-541. We acknowledge the support from INAF and Ministero dell' Istruzione, dell' Universita' e della Ricerca (MIUR) in the form of the grant "Premiale VLT 2012" and the grant "The Chemical and Dynamical Evolution of the Milky Way and Local Group Galaxies" (prot. 2010LY5N2T). The results presented here benefit from discussions held during the Gaia-ESO workshops and conferences supported by the ESF (European Science Foundation) through the GREAT Research Network Programme. This research has made use of the SIMBAD database, operated at CDS, Strasbourg, France, NASA's Astrophysics Data System, and the WEBDA database, operated at the Department of Theoretical Physics and Astrophysics of Masaryk University.
\end{acknowledgements}

\bibliographystyle{aa} 
\bibliography{../smiljanic}

\end{document}